%% file: main.tex
\documentclass[twocolumn,twocolappendix]{aastex63}
\usepackage{graphicx}
\usepackage{amsmath}
\usepackage{multirow}
\usepackage{chngcntr}

\newcommand{\xj}[1]{{#1}} 
\newcommand{\update}[1]{ { {#1}} }
\newcommand{\ha}{H$\alpha$}
\newcommand{\lya}{Ly$\alpha$}
\newcommand{\fesclya}{$f_{\rm esc,Ly\alpha}$}
\newcommand{\fesclyc}{$f_{\rm esc,LyC}$}
\newcommand{\fesclycrel}{$f_{\rm esc,LyC}^{\rm rel}$}
\newcommand{\fesclyarel}{$f_{\rm esc,Ly\alpha}^{\rm rel}$}
\newcommand{\xiion}{$\xi_{\rm ion}$}
\newcommand{\xiionzero}{$\xi_{\rm ion,0}$}
\newcommand{\Ms}{$M_\star$}
\newcommand{\LUVint}{$L_{\rm UV,int}$}
\newcommand{\LUVatt}{$L_{\rm UV,att}$}
\newcommand{\MUVint}{$M_{\rm UV,int}$}
\newcommand{\MUVatt}{$M_{\rm UV,att}$}
\newcommand{\betaobs}{$\beta_{\rm obs}$}
\newcommand{\SDsSFR}{$\Sigma_{\rm sSFR}$}
\newcommand{\EBV}{E($B-V$)}

\newcommand{\nHAE}{165}
\newcommand{\fesclyaCal}{0.090$\pm$0.006}
\newcommand{\fesclyaSMC}{0.121$\pm$0.009}

\defcitealias{Calzetti2000}{Calz}

\received{xxx}
\revised{xxx}
\accepted{xxx}

\submitjournal{ApJ}

\shorttitle{\lya\ escape at $z=5-6$}
\shortauthors{Lin et al.}

\graphicspath{{./}{figures/}}

\begin{document}

\title{ Quantifying the escape of \lya\ at $z\approx 5-6$: a census of  \lya\ escape fraction with \ha\ emitting galaxies spectroscopically confirmed by JWST and VLT/MUSE} 

\author[0000-0001-6052-4234]{Xiaojing Lin}
\affiliation{Department of Astronomy, Tsinghua University, Beijing 100084, China}
\email{linxj21@mails.tsinghua.edu.cn, zcai@mail.tsinghua.edu.cn}
\affil{Steward Observatory, University of Arizona, 933 N Cherry Ave, Tucson, AZ 85721, USA}

\author[0000-0001-8467-6478]{Zheng Cai}
\affiliation{Department of Astronomy, Tsinghua University, Beijing 100084, China}

\author[0000-0003-0111-8249]{Yunjing Wu}
\affiliation{Department of Astronomy, Tsinghua University, Beijing 100084, China}
\affil{Steward Observatory, University of Arizona, 933 N Cherry Ave, Tucson, AZ 85721, USA}

\author[0000-0001-5951-459X]{Zihao Li}
\affiliation{Department of Astronomy, Tsinghua University, Beijing 100084, China}

\author[0000-0002-4622-6617]{Fengwu Sun}
\affiliation{Steward Observatory, University of Arizona, 933 N Cherry Ave, Tucson, AZ 85721, USA}

\author[0000-0003-3310-0131]{Xiaohui Fan}
\affil{Steward Observatory, University of Arizona, 933 N Cherry Ave, Tucson, AZ 85721, USA}

\author[0000-0002-2178-5471]{Zuyi Chen}
\affil{Steward Observatory, University of Arizona, 933 N Cherry Ave, Tucson, AZ 85721, USA}

\author[0000-0001-6251-649X]{Mingyu Li}
\affiliation{Department of Astronomy, Tsinghua University, Beijing 100084, China}

\author[0000-0002-1620-0897]{Fuyan Bian}
\affiliation{European Southern Observatory, Alonso de C\'{o}rdova 3107, Casilla 19001, Vitacura, Santiago 19, Chile} 

\author[0000-0001-9442-1217]{Yuanhang Ning}
\affiliation{Department of Astronomy, Tsinghua University, Beijing 100084, China}

\author{Linhua Jiang}
\affiliation{Kavli Institute for Astronomy and Astrophysics, Peking University, Beijing 100871, China}
\affiliation{Department of Astronomy, School of Physics, Peking University, Beijing 100871, China}


\author{Gustavo Bruzual}\affiliation{Institute of Radio Astronomy and Astrophysics, National Autonomous University of Mexico, San José de la Huerta
58089 Morelia, Michoacán, México}

\author[0000-0003-3458-2275]{Stephane Charlot}\affiliation{Sorbonne Universit\'e, CNRS, UMR7095, Institut d'Astrophysique de Paris, F-75014, Paris, France}

\author[0000-0002-7636-0534]{Jacopo Chevallard}
\affiliation{Department of Physics, University of Oxford, Denys Wilkinson Building, Keble Road, Oxford OX1 3RH, UK}

\begin{abstract}
The James Webb Space Telescope (JWST) provides an unprecedented opportunity for unbiased surveys of \ha-emitting galaxies at $z>4$ with the NIRCam’s wide-field slitless spectroscopy (WFSS). In this work, we present a census of Ly$\alpha$ escape fraction (\fesclya) of \nHAE\ star-forming galaxies at $z=4.9-6.3$, utilizing their \ha\ emission directly measured from FRESCO NIRCam/WFSS data. We search for  \lya\ emission of each H$\alpha$-emitting galaxy in the VLT/MUSE data. The overall \fesclya\  measured by stacking is \fesclyaCal. We find that \fesclya\ displays a strong dependence on the observed UV slope (\betaobs) and \EBV, such that the bluest galaxies (\betaobs$\sim-2.5$) have the largest escape fractions (\fesclya$\approx0.6$), indicative of the crucial role of dust and gas in modulating the escape of \lya\ photons. \fesclya\ is less well related to other parameters, including the UV luminosity and stellar mass, and the variation in \fesclya\ with them can be explained by their underlying coupling with \EBV\  or \betaobs. Our results suggest a tentative decline in \fesclya\  at $z\gtrsim 5$, implying increasing  intergalactic medium attenuation towards higher redshift. Furthermore, the dependence of \fesclya\ on \betaobs\ is proportional to that of the ionizing photon escape fraction (\fesclyc), indicating that the escape of \lya\ and ionizing photon may be regulated by similar physical processes. With \fesclya\ as a proxy to \fesclyc, we infer that UV-faint  ($M_{\rm{UV}}>-16$) galaxies contribute $>70\%$ of the total ionizing emissivity  at $z=5-6$. If these relations hold during the epoch of reionization, UV-faint galaxies can contribute the majority of UV photon budget to reionize the Universe.

\end{abstract}

\keywords{\lya\ escape --- reionziation sources --- Lyman continuum --- high-redshift}

\input{01introduction}
\bigskip

\input{02data_and_sample}

\bigskip

\input{03measurement}

\bigskip

\input{04result}

\bigskip

\input{05discussion}
\bigskip

\input{06summary}
\bigskip

\input{99DataAvail}
\bigskip

\input{99other}

\bigskip

\appendix
\input{99appendix}
\bigskip

\bibliography{main}{}
\bibliographystyle{aasjournal}

\end{document}

%% file: 01introduction.tex
\section{Introduction}

The Epoch of Reionization (EoR) represents a 
phase transition 
when the intergalactic medium (IGM) evolved rapidly from mostly neutral to highly ionized \citep{Fan2006,Robertson2022}. Although it is generally acknowledged that EoR is completed by $z\sim 5-6$ \citep[e.g.,][]{Kulkarni2019,Zhu2021,Jin2023}, its onset and major contributors remain elusive. 
\xj{Recent measurements on quasar luminosity functions at $z\sim6$ suggest that the quasar population only contributes a negligible fraction ($\lesssim 7-10\%$) of total photons to keep the universe ionized \citep{Matsuoka2018, Jiang2022}. Young star-forming galaxies are thought to be the main producers of ionizing photons for reionization. However,} one of the biggest uncertainties is whether a vast number of faint galaxies \citep[e.g.,][]{Finkelstein2019} or a small population of bright oligarchs \citep[e.g.,][]{Naidu2020} dominate the cosmic ionizing emissivity during the EoR. To address these questions, extensive studies in the past decades focused on three key quantities observationally: the ionizing photon production efficiency \xiion, the UV luminosity density $\rho_{\rm UV}$, and the escape fraction of Lyman-continuum (LyC) photons \fesclyc\ \citep{Robertson2013,Bouwens2016}.

Among the above three 
elements, the biggest uncertainty is  
from the estimates of \fesclyc\ at high redshifts \citep{Vanzella2010,Bian2020,Naidu2022}. Direct observing the leakage of LyC photons at $z \gtrsim4$ \citep{Steidel2018} is difficult due to the high opacity of the IGM. Therefore,  many works studied
low-redshift galaxies with similar properties \citep[e.g.,][]{Leitherer2016,Flury2022,Saldana-Lopez2022} or indirect tracers of \fesclyc\ \citep[e.g.,][]{Flury2022b,Xu2022,Choustikov2023}, 
although it is still unclear whether low-redshift analogs can be good representatives of the main contributors to reionization 
\cite[e.g.,][]{Katz2020,Schaerer2022}, and whether those indirect indicators of \fesclyc\ derived from very small sample could be generalized  \cite[e.g.,][]{Katz2022,Saxena2022}. The typically required \fesclyc\ for reionization to end at $z\sim 5-6$ is approximately $\gtrsim$5\% \citep{Robertson2015,Finkelstein2019}, and recent work have constrained the average \fesclyc\ to be $\lesssim$ 10\% through various indirect methods \citep{Steidel2018,Pahl2021,Saldana-Lopez2023}.

To directly constrain the \fesclyc\ for galaxies at $z\gtrsim4$, 
one of the most promising 
ways is to study the escape of \lya. 
Numerous studies have shown 
a strong connection between the escape of ionizing photons and 
\lya\  photon \citep{Chisholm2018,Marchi2018,Maji2022}. Their escapes are both determined by the 
distribution and geometry of 
neutral gas in the interstellar medium (ISM), and facilitated by optically-thin channels within the gas reservoir \citep{Dijkstra2016,Gazagnes2020,Lin2023}.  These ionized channels could be carved by feedback from massive young stars, which also modulate the production of LyC and \lya\ photons \citep{Vanzella2022,Kim2023}. 

A number of previous efforts have been made to probe the escape of 
Ly$\alpha$ (\fesclya) which could be derived by comparing 
Ly$\alpha$ and \ha\ emissions. 
Before the launch of James Webb Space Telescope (JWST), \ha\ of galaxies at $z\gtrsim5$ could be estimated only through the \textit{Spitzer} broadband color \citep[e.g.,][]{Faisst2019, Lam2019,Stefanon2022}. Hence the measurements of \fesclya\ in the pre-launch era suffer 
from large modeling and photometric uncertainties. Improved estimates of the \ha\ flux excess have been done recently using JWST narrow band observations \citep{Ning2023, Simmonds2023}. Moreover, the 
spectroscopic capability of JWST in the infrared regime 
 enables us to detect 
and accurately measure \ha\ emission lines in faint, high-redshift galaxies. The JWST Near Infrared Spectrograph (NIRSpec) can capture both the \lya\ and \ha\ emission of high-redshift galaxies simultaneously and thus quantify \fesclya \citep{Saxena2023, Chen2023, Napolitano2024}, but a census of \fesclya\ by NIRSpec is difficult due to the limited sample size it can observe and the trade-off between high sensitivity and high spectral resolution.
The wide-field slitless spectroscopy of JWST/NIRCam \citep[WFSS;][]{Greene2016, Rieke2023}, instead, offers a unique capability to map the spectra of all sources within its field of view (FoV) \citep{Sun2022a,Sun2022b}, ideal to build a  large sample of emission line galaxies through blind spectroscopic surveys. 
Combing the \ha\ emission lines captured by WFSS with the corresponding \lya\ lines observed by ground-based facilities, we are able to,  for the first time, conduct a direct census of \fesclya\ with an unbiased sample limited by \ha\ flux.

In this work, we utilize JWST/NIRCam WFSS around $ 4\mu m$ to detect \ha\ emission over the redshift range of $z\approx 4.9-6.3$, when the Universe was experiencing rapid transition and galaxies were undergoing substantial mass, dust and chemical content assembly \citep{Madau2014, Davidzon2017, Faisst2019}. The ground-based VLT/MUSE \citep{Bacon2010}, working in the full optical domain, perfectly covers their \lya\ so that we can obtain more precise measurements of \fesclya.

The paper is organized as follows. In \S\ref{sec:data and sample} we outline the photometric and spectroscopic data used and sample selection based  on \ha\ emission lines. We describe the measurements of galaxy properties, search for \lya\ lines, and methods to calculate \fesclya\ in \S\ref{sec:measurement}. In \S\ref{sec:result} we present the distribution of \fesclya, its dependence on galaxy properties and redshift evolution. We then investigate the connection between LyC and \lya\ escape in \S\ref{sec:discussion}, and discuss the implication for reionization. The final summary is presented in \S\ref{sec:summary}. We show more details about the analysis and address potential systematics in the Appendix. Throughout this work, a flat $\Lambda$CDM cosmology is assumed, with $\rm H_0 = 70~km~s^{-1}~Mpc^{-1} $, $\Omega_{\Lambda,0} = 0.7$ and  $\Omega_{m,0}=0.3$. 

%% file: 02data_and_sample.tex
\section{Data and sample}\label{sec:data and sample}

We describe the dataset that we use in this work and the construction of our sample in this section. We introduce all the photometric and spectroscopic data, including JWST, HST and VLT/MUSE, in \S\ref{sec:phot-spec-data}. 
The selection of \ha\ emitters is presented in \S\ref{sec:ha sample}.  

\subsection{Photometric and Spectroscopic Data}\label{sec:phot-spec-data}
\subsubsection{JWST/NIRCAM Imaging and Slitless Grism in the GOODS-S field}

The JWST FRESCO (``First Reionization Epoch Spectroscopically Complete Observations'') Survey  \citep[GO-1895; PI Oesch]{Oesch2023},  covers 62 arcmin$^2$ in each of the extragalactic legacy GOODS/CANDELS fields with deep NIRCam imaging and wide field slitless spectroscopic (WFSS) observations.  In GOODS-South field, FRESCO obtains $4\times2$ pointings with 8$\times$7043 s exposures using the F444W grism, accompanied by $8\times 3522$ s exposures of F210M imaging and $11\times 4456$ s of F182M imaging. Additional $8\times934$ s exposures of F444W imaging are taken to associate the sliteless spectra with individual objects.  

The imaging data were firstly reduced up to \texttt{stage2} using the standard JWST pipeline v1.7.2\footnote{\url{https://github.com/spacetelescope/jwst}} and the calibration reference files ``jwst\_1014.pmap''.  The calibrated single exposures (\texttt{*\_cal.fits}) were further processed by \textsc{Grizli}\footnote{\url{https://github.com/gbrammer/grizli/}}. \textsc{Grizli} mitigated 1/f noise, alleviated ``snowball'' artifacts from cosmic rays, and converted the world coordinate
system (WCS) information in the headers of each exposure to SIP (Simple Imaging Polynomial) format so that exposures can be drizzled and
combined with \textsc{Astrodrizzle}. The WCS of each exposure image was registered using GAIA DR3 \citep{Gaia2023} and the images were finally drizzled with a pixel scale of 0.031\arcsec using \texttt{pixfrac=0.8}. An additional background subtraction using \textsc{photutils.Background2D} was performed on the final mosaics after masking bright sources.

For the WFSS data, after the standard \texttt{stage1} steps, flat-fielding was performed using the flat of direct imaging since the WFSS flat reference files are not available yet in CRDS. 1/f noise is removed only along columns as the spectra are dispersed along rows. 
The WCS for each individual exposure was assigned using \texttt{CALWEBB.assign\_wcs}. There were astrometric offsets between this assigned WCS and that of the fully-reduced F444W images. To associate the WFSS spectra and individual galaxies accurately, we calculated the offsets using accompanied short-wavelength exposures (i.e., the F182M and F210M imaging taken simultaneously with the WFSS) and corrected them for each WFSS exposure. These pre-processed files, as well as a source catalog based on the final F444W mosaic, were provided as inputs for \textsc{Grizli} for spectral extraction using the updated sensitivity functions and tracing models as described in \cite{Sun2022b}\footnote{\url{https://github.com/fengwusun/nircam_grism}}.  To facilitate emission line searching, we performed running median filtering along the dispersion direction on single calibrated grism exposures as described in \cite{Kashino2023},  and extracted continuum-removed 1D spectra aiming for emitter search only.

\subsubsection{HST Imaging in GOODS-S field}
The high-level products of the Hubble Legacy Fields\footnote{\url{http://archive.stsci.edu/hlsps/hlf}} \citep[HLF;][]{Whitaker2019} in GOODS-S include images from ultraviolet to infrared bands, observed over the past 18 years, with 7491 individual exposures and 6.4 million seconds in total. All the HLF image mosaics are tied to an absolute GAIA DR2 \citep{Gaia2018} reference frame. In this work, we use four optical images in the ACS/WFC F435W, F606W, F775W, F814W bands at 30mas/pixel, and four infrared images in the WFC3/IR F105W, F125W, F140W and F160W bands at 60mas/pixel.

\subsubsection{Source Extraction and Photometry}\label{sec:source_phot}
We first resampled all HST images onto a fixed grid with a pixel scale of 0.031\arcsec. For HST mosaics we constructed empirical PSFs by stacking stars with good SNRs in the fields. For JWST mosaics,  \xj{the centers of
most bright stars are saturated especially in the F182M and F210M images, so it is hard to build empirical PSFs that can accurately reproduce the complicated PSF shapes with these limited numbers of stars in the final mosaics. Therefore,} we adopted the modeled PSFs provided by \textsc{WebbPSF}\footnote{\url{https://github.com/spacetelescope/webbpsf}}. We generate point-spread function (PSF) matching kernels to match the PSFs of HST WFC3/IR F160W or JWST/NIRCam F444W, using \textsc{Photutils \footnote{\url{https://photutils.readthedocs.io/en/stable/psf_matching.html}} }, with  high-frequency noise filters in \cite{Aniano2011}. All HST ACS/WFC mosaics (F435W, F606W, F775W, F814W), JWST/NIRCam  F182M, F210M mosaics were convolved to match the point-spread function (PSF) of JWST/NIRCam F444W with corresponding kernels; all HST WFC3/IR mosaics (F105W, F125W, F140W) were matched to the PSF of F160W. 

We ran \textsc{Source Extractor} \citep{SExtractor} in dual mode for each band, using the F444W mosaics as the detection image. We started with Kron aperture photometry with Kron factors $k = 1.2$. We measured the flux uncertainty for each object by randomly positioning the same Kron aperture in empty regions nearby. The measured flux and uncertainties were further corrected to total values  by multiplying the ratio of flux inside the $k = 2.5$ v.s. $k = 1.2$ Kron apertures on the detection image. Finally, we correct for flux outside the $k=2.5$ radii using the PSF of F444W. Note that for HST WFCS/IR bands we homogenized the images to the PSF of F160W, 
slightly larger than that of JWST/NIRCam F444W, thus we enlarged the aperture size on these HST IR mosaics by a factor of 1.2, which ensures that all the flux from objects would be included as tested on the F160W PSF.

\subsubsection{VLT/MUSE in GOODS-S field}

The Multi Unit Spectroscopic Explorer (MUSE) is an integral-field spectrograph in the the optical wavelength range on the Very Large Telescope (VLT). MUSE has a field of view (FoV) of one arcmin$^2$, a spatial sampling of 0.2\arcsec, and a spectral range from 4750\AA\ to 9350\AA\ with a mean resolution of 3000.  Its high sensitivity to emission lines in the full optical domain makes it an ideal instrument to detect high-redshift \lya\ emitters (LAEs) over $z=2.9-6.7$.

The GOODS-S field is covered by the legacy MUSE-Wide survey \citep{Urrutia2019} and MUSE Hubble Ultra Deep Field (UDF) survey \citep{Bacon2023}. MUSE-Wide comprises 60$\times$1 arcmin$^2$ pointings, each with a depth of 1 hour. The MUSE UDF survey includes $3\times3$ arcmin$^2$ mosaic of nine 10-hour depth fields (MOSAIC),  a $1\times1$ arcmin$^2$ 31-hour depth field (UDF-10), and the circular MUSE eXtremly
Deep Field (MXDF). The MXDF covers a radius of 41\arcsec and 31\arcsec for respectively 10+
and 100+ hours of depth, reaching a final maximum depth of 141 hours, and is the deepest optical spectroscopic survey so far. In Figure \ref{fig:HAE_distribution} we show the footprints of FRESCO, MUSE-Wide and MUSE UDF. The overlap region between FRESCO and the two MUSE surveys is about 50 arcmin$^{2}$, including $\sim43$ arcmin$^{2}$ of the MUSE-Wide, $\sim9$ arcmin$^{2}$ of the 10-hour MOSAIC and the whole UDF-10 and MXDF.

We adopt the publicly released DR1\footnote{\url{https://musewide.aip.de/}} data for MUSE-Wide and \textsc{AMUSED}\footnote{\url{https://amused.univ-lyon1.fr/project/UDF/HUDF/}} data for MUSE UDF.  For MUSE-Wide, DR1 only releases 44 of the 60 datacubes, so we reduced the remaining 16 datacubes 
with our custom pipeline. This includes the standard pipeline\footnote{\url{https://www.eso.org/sci/software/pipelines/muse/}} \texttt{v2.8.4} \citep{Weilbacher2020} and improved the sky subtraction using \textsc{ZAP}\footnote{\url{https://github.com/musevlt/zap}} \citep{Soto2016}. 
We aligned all the datacubes to the Hubble ACS astrometry as 
done by the MUSE-Wide team. The offset between the HST astrometry  and the GAIA DR2, 
adopted in WFSS and imaging, is $\Delta \rm{Ra} = +0.094\pm
0.042\arcsec$ and $\Delta \rm{Dec} = -0.26 \pm 0.10 \arcsec$ \citep{Bacon2023}.  

\subsection{Selection of \ha\  emitters at $z=5-6$}\label{sec:ha sample}

\subsubsection{Selecting high-redshift emitters both photometrically and spectroscopically}\label{sec:selection}


The  F444W WFSS of JWST/NIRCam is used to conduct unbiased surveys of \ha\ emitters (HAEs) at $z=5-6$.  We performed emission line searching  on the 1D continuum-removed spectra first. Among all the emitter candidates, we identified possible high-redshift emitter candidates by applying the following cuts:

\begin{itemize}
    \item[(1)] 1D emission line with SNR$\geq$4,
    \item[(2)]  SNR$<2$ in F435W,
    \item[(3)] F606W-F105W$>$1 and F606W-F182M$>$1
    \item[(4)] F182M-F444W$>$-0.1
\end{itemize}

The second criterion requires a complete 912 \AA\  dropout for $z>4.9$ galaxies at F435W. The third criterion describes the \lya\ break features; we require at least 1 mag drop in F606W that is blueward the wavelength of \lya.  
The fourth criterion is the F444W excess due to \ha\ emission lines when considering possible Balmer jumps. This is the minimum excess for  $z=4.9-6.7$ mock galaxies in \texttt{JAGUAR} \citep{Williams2018} with \ha\ flux larger than $1.6\times10^{-18}$ erg s$^{-1}$ cm$^{-2}$ (4$\sigma$ depth of the FRESCO F444W grism data according to \cite{Oesch2023}).   We also obtained photometric redshifts for each source using \textsc{EAZY} \citep{EAZY} with the \texttt{corr\_sfhz\_13} templates, which contain redshift-dependent star formation histories. We excluded objects with robust photo-$z$, i.e. $P(z_p>4.5)=0$, as low-redshift objects.  We finally constructed a $z>4.9$ emitter catalog by careful visual inspection, during which we removed artifacts on imaging and grism data, false detection by the search algorithms, and misidentified lines from contamination of other sources. 

\subsubsection{HAE sample}\label{sec:final_ha_sample}

The high-redshift emitter catalog described above comprises all $z>4.9$ candidates, including $z>7$ [OIII] emitters. Bright [OIII] emitters are easily identified by their doublets at restframe 4959 and 5007 \AA. However, as the [OIII] 4959\AA\  lines are $\sim 3$ times fainter than the 5007\AA, and 
H$\beta$ can be even fainter,  in many cases we can only detect a single 5007\AA\ emission while the 4959\AA\ lines remain undetected (see Appendix \ref{sec:sample_selection_detail} for more details). Since our science goal is more sensitive to the sample purity 
than completeness, we exclude all possible [OIII] candidates based on both photometry and spectroscopy. We firstly assumed the brightest single emission line in the 1D spectra to be 5007\AA\ lines and force-fitted 4959 lines. This would effectively identify [OIII] emitters with marginally-detected 4959 lines.  We then further extend the [OIII] 
sample  by 
photometric cuts as follows:
\begin{itemize}
    \item SNR$<2$ in both F606W and F775W.
    \item F606W-F115W$>$1.7 and F814W-F115W$>$1.7
    \item F115W-F210M$<$1.0 and F115W-F210M$<$1.0
    \item F814W-F115W$>$(F115W-F210M)+1.5
\end{itemize}
Similar criteria have been applied in \cite{Endsley2022} to select $z>6.5$ emitters with high [OIII] equivalent width. Note that we do not require F814W to be completely undetected, allowing for \lya\ transmission spikes within its wide wavelength coverage\citep{Kakiichi2018}.

We find 
74 \textit{possible} [OIII] candidates. 29 of the 74 [OIII] candidates reside in the JADES-DEEP footprint \citep{Rieke2023,Eisenstein2023}. Among the 29 sources,  21 sources have photo-$z>7$  based on multiple bands of JADES imaging, with [OIII] emission in F444W; 8 of them have photo-$z$ of $5-6$ with \ha\ in F444W. We argue that our criteria for [OIII] could, in most cases, successfully remove  $z>7$ [OIII] emitters and thus keep an \ha\ sample with high purity. 

\update{Finally we obtain a sample of 222 HAEs. We identify blended clumpy galaxies with irregular morphology and multiple cores as single objects. These clumpy HAEs have \ha\ emission lines that cannot be distinguished on the 2D grism spectra, and some of them have unsolved \lya\ emission captured by MUSE, even though \textsc{Sextractor} may label them as different individuals. During the extraction of grism spectra, we use a modified segmentation map as the input of \textsc{grizli}, in which clumpy segments are combined. In our final sample, any two `single' objects are required to be separated from each other 
by at least 3.5 kpc (i.e., 0.6 \arcsec at $z=5.5$). } 
We measure the total \ha\ flux on 1D extracted spectra by fitting a `Gaussian + constant' model, where the constant is used to depict the local continuum level around \ha. 

Moreover, we label sources having neighbors within 25 kpc (about $4\arcsec$ at $z=5-6$) and $\pm 500$ km s$^{-1}$.  Due to the significant disparity between the spatial resolutions of MUSE and JWST, we will exclude these galaxies and their close neighbors in the following analysis, since (1) \lya\ around individual galaxies can be extended to  $3\arcsec$ \citep{Wisotzki2016} and will boost \lya\ in the overlapping regions of close pairs, introducing systematics during the stack of \lya\ 
as described in \S\ref{sec:lya stack}, and (2) for very close pairs it is difficult to obtain accurate \ha\ flux due to the overlap issue along the dispersion direction of grism.  
Among all the 222 HAEs,  21 of them have close neighbors. Three of the 222 HAEs are bright point sources with hints as being AGNs. There are 184 HAEs that reside in the MUSE footprints, with 17 having close neighbors and two as potential AGNs.  Finally, \nHAE\ HAEs are selected for the following studies. The spatial and redshift distributions of our selected sample are shown in Figure \ref{fig:HAE_distribution}. Detailed properties of these HAEs are presented in Appendix \ref{sec:sample_selection_detail}, which demonstrate that they are typical main-sequence galaxies in this redshift range.

\begin{figure*}[htbp!]
\centering
\includegraphics[width=\textwidth]{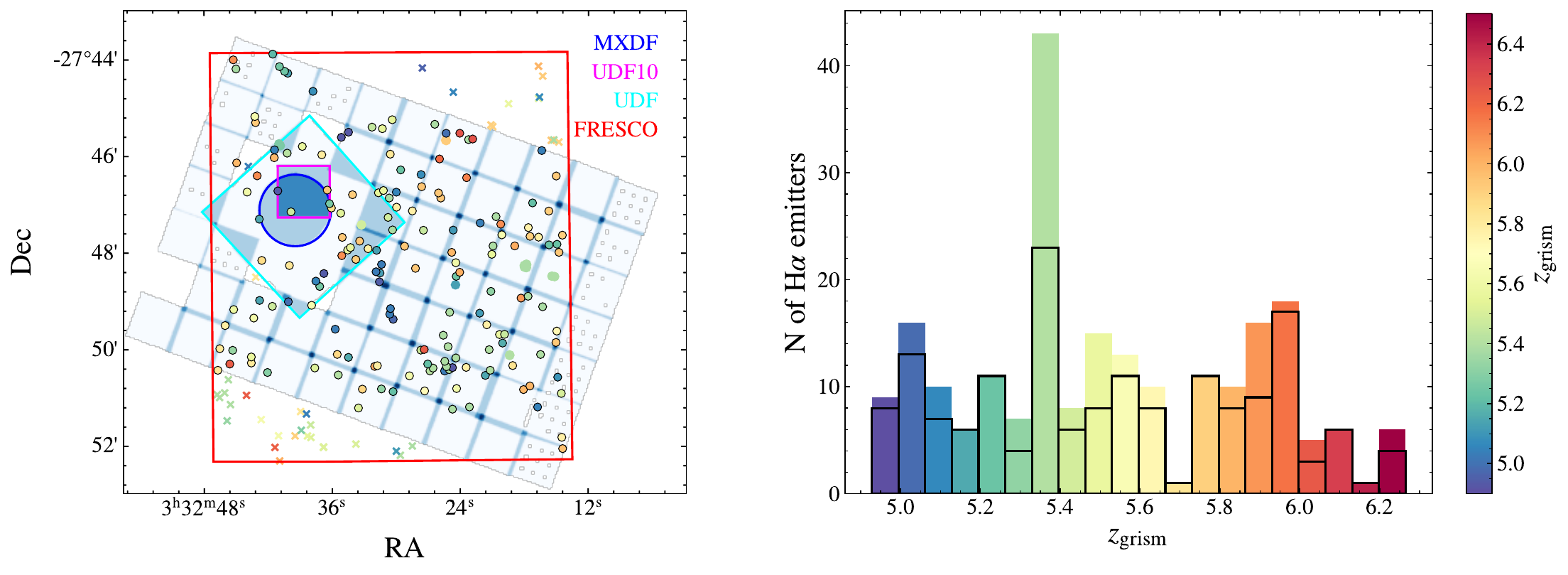}
\caption{The spatial and redshift distribution of \ha\ emitters. \textit{Left}: the distribution of \ha\ sample in the footprints of multi-band legacy data in GOODS-S. The blue shaded region is covered by MUSE observations, {including the MUSE-Wide \citep{Urrutia2019} and MUSE Hubble UDF survey \citep{Bacon2023}},  color-coded by the number of overlapping datacubes (darker color indicates more datacubes). The cyan, blue, and magenta frames represent the footprints of three fields in MUSE Hubble UDF survey: {the 10-hour UDF, 31-hour UDF-10, and the $>100$-hour MXDF}; the red rectangle shows the FoV of FRESCO JWST/NIRCam F444W imaging. Color-coded markers indicate the position of our \ha\ emitters. The color indicates their redshift in exactly the same scale as that on the right panel. An overdensity at $z\approx 5.3-5.4$ is obvious, which has been reported in \cite{Helton2023}. The crosses denote \ha\ emitters not covered by MUSE (41 in total); dots are within MUSE coverage (197 in total) and dots with black edges are selected sources we use for analysis (157 in total). \textit{Right}: the redshift distribution of our \ha\ emitters. The distribution of selected \ha\ sources is wrapped with black edges. }
\label{fig:HAE_distribution}
\end{figure*}

%% file: 03measurement.tex
\section{Measurements}\label{sec:measurement}
We describe the methods we use to quantify the properties of our HAE sample in \S\ref{sec:SED}.  The search and measurements of \lya\ emission using VLT/MUSE data are outlined in \S\ref{sec:MUSE_lya}. The calculation of \fesclya\ for individual galaxies and through stack analysis is presented in \S\ref{sec:measure_fesclya}.

\subsection{SED modelling and galaxy properties}\label{sec:SED}
We infer the physical properties of the \nHAE\  $z\approx 5-6$ HAEs by SED modeling with the Bayesian code \textsc{Beagle} \citep{Chevallard2106}, using the measured photometry (see \S\ref{sec:source_phot}) and \ha\ line flux as inputs. \xj{We do not use the \lya\ flux as inputs because modeling Ly$\alpha$ and H$\alpha$ simultaneously requires correctly taking the Ly$\alpha$ escape process into account and current SED modeling codes cannot do it.} We assume a constant star formation history and the \cite{Chabrier2003} initial mass function (IMF) with an upper limit of 100${\rm M}_\odot$. We adopt flat priors for metallicity, $0.01<{ Z}/{\rm Z}_\odot<0.2$, and ionization parameter for nebular emission  \citep{Gutkin2016} in log-space, $-3<\log{U}<-1$, which is the dimensionless ratio of the number density of H-ionizing photons to that of hydrogen. For dust attenuation, we adopt two attenuation curves: the Small Magellanic Cloud (SMC) dust law \citep[][hereafter SMC]{Pei1992} and the \cite{Calzetti2000} law (hereafter \citetalias{Calzetti2000}) with $R_V$ (ratio of total to selective extinction in V band) being 2.74  and 4.05 respectively. We set the optical depth in the V band varying in the range of 0 -- 0.5 in log space. Although a large variety of dust attenuation curves for high-redshift galaxies have been used in the literature \citep{Reddy2015,Scoville2015,Reddy2018}, these two are commonly suggested to be applicable and utilized broadly in high-redshift studies \citep[e.g.,][]{Bouwens2016,Shivaei2018,Faisst2019}. \update{In the following analysis, we 
adopt the \citetalias{Calzetti2000} as our fiducial dust attenuation law, but we also consider the SMC law for comparison.}

The \textsc{Beagle} SED fitting results can provide the stellar mass (\Ms), the effective optical depth at the V band ($\tau_V$), the attenuated UV luminosity \LUVatt\  and the intrinsic UV luminosity \LUVint\ directly. The UV luminosity is defined at restframe 1600~\AA, and the optical depth could be converted to the color excess \EBV\  with corresponding attenuation curve. We measure the observed UV slope ($\beta_{\rm obs}$, not corrected for dust) by fitting the stellar continua of the best-fit SED models with a power-law function ($f_\lambda=\lambda^{\beta_{\rm obs}}$),  spanning from 1260~\AA\ to 2500 \AA.  \xj{The SED-derived UV luminosity and UV slopes are statistically consistent with the fitting results using the photometry directly.}  We estimate the specific star formation rate (sSFR) for each galaxy with their intrinsic \ha\ luminosity and SED-derived \Ms. The SFR of each galaxy is converted based on \cite{Kennicutt1998} specialized for the \cite{Chabrier2003} IMF: 
\begin{equation}\label{eq:SFR_Ha}
\log{\rm SFR[ M_\odot~ yr^{-1}]} = \log L_{\rm H\alpha}{[\rm erg~ s^{-1}]} - 41.35. 
\end{equation}
Note that the conversion factors between the \ha\ luminosity and SFR for high-redshift galaxies vary among different stellar population models and IMFs, and are highly dependent on the ionizing photon production efficiency \xiion\  \citep[e.g.,][]{Theios2019}.  The conversion factor of subsolar-metallicity galaxies is about 2.5 lower than the canonical conversion factors of low-redshift galaxies \citep[e.g.,][]{Reddy2018, Theios2019, Shapley2023}. Nevertheless, as we focus on the relative relation between \fesclya\ and SFR in the following analysis (see \S\ref{sec:fesclya_galaxy_properties}), Equation \ref{eq:SFR_Ha} is qualitatively a good proxy to the SFR of our \ha\ emitting galaxies. 

To derive the sSFR surface density (\SDsSFR) , we fit the F444W surface brightness distribution, i.e. the approximate \ha\ surface brightness, for each galaxy using single or multiple Sérsic models. The effective radius ($r_{\rm eff}$), which is the radius of the circular aperture including half of the total flux from the galaxy, is measured in the corresponding compound modeled image after PSF deconvolution\footnote{\url{https://users.obs.carnegiescience.edu/peng/work/galfit/TFAQ.html\#effective_radius}}. We do not adopt $r_{\rm eff}$ in simple single Sérsic modeling due to the clumpy morphology as well as the highly coupled parameters during modeling. Then we calculate \SDsSFR\ as 
\begin{equation}
\centering
    \Sigma_{\rm sSFR}={\rm sSFR}/2\pi r_{\rm eff}^2.
\end{equation}

We further measure the ionizing efficiency \xiion, i.e the production rate of Lyman-continuum (LyC) ionizing photons per unit non-ionizing UV continuum luminosity. \xiion\ can be expressed in terms of $N\left(\mathrm{H}^0\right)$, the production rate of LyC
photons, and \LUVint\ as
\begin{equation}
\xi_{\text {ion }}=\frac{N\left(\mathrm{H}^0\right)}{L_{\mathrm{UV, int}}}\left[\mathrm{s}^{-1} / \mathrm{erg} \mathrm{s}^{-1} \mathrm{~Hz}^{-1}\right]
\end{equation}
where 
\begin{equation}
L_{\mathrm{H \alpha,int}}=1.36 \times 10^{-12}\left(1-f_{\rm esc, LyC}\right) N\left(\mathrm{H}^0\right).
\label{eq:Lha_fesc}
\end{equation}
\fesclyc\ is the escape fraction of LyC photons from galaxies. Our result, as shown in \S\ref{sec:discussion}, shows that \fesclyc\ has negligible impact to the estimate of \xiion, so we assume \fesclyc\ as zero in Equation \ref{eq:Lha_fesc}. Therefore, \xiion\ we adopt is actually the production efficiency of ionizing photons failing to escape from the galaxy.

\begin{figure}
    \centering
    \includegraphics[width=\columnwidth]{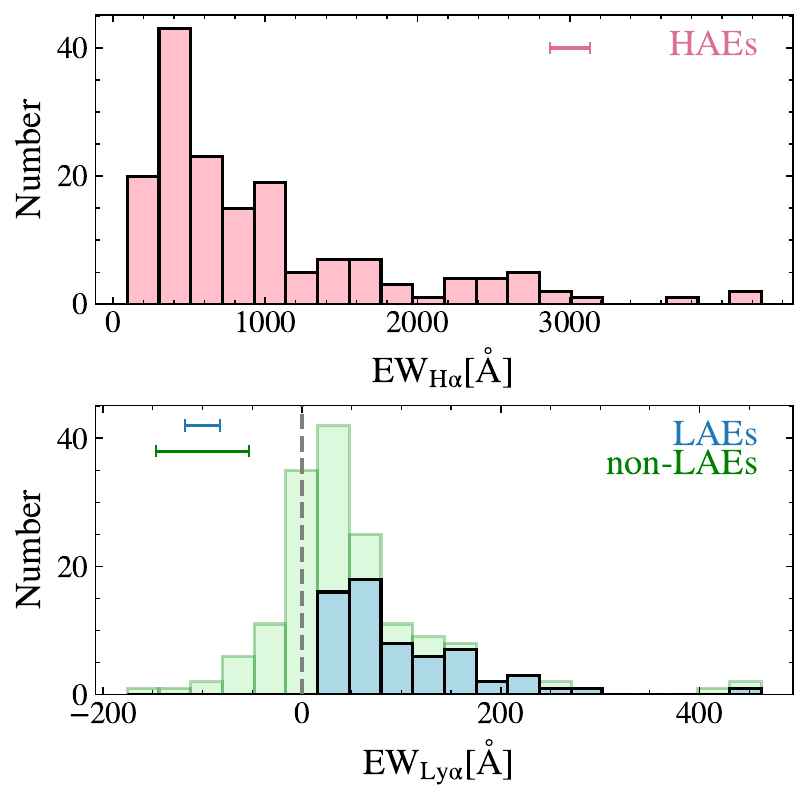}
    \caption{The distribution of EW$_{\rm H\alpha}$  and  EW$_{\rm Ly\alpha}$ of the \nHAE\  selected HAEs in the overlapping region of FRESCO and MUSE footprints. The EWs are calculated following Equation \ref{eq:EW_cal} with $f_{\rm cont}$ from the SED-modeled stellar continuum assuming the \citetalias{Calzetti2000} dust attenuation law. We label the mean uncertainty of EW$_{\rm H\alpha}$ (pink in the left panel) and EW$_{\rm Ly\alpha}$ (blue for LAEs and green for non-LAEs in the right panel) as horizontal lines, respectively.  For non-LAEs, we note that the uncertainty of forced measured EW$_{\rm Ly\alpha}$ reaches an average of 163\AA\ when  EW$_{\rm Ly\alpha}>100\AA$, though the mean uncertainty for non-LAEs is about 47\AA. We mark the EW$_{\rm Ly\alpha}=0$ as the gray dashed line.  }
    \label{fig:EW_distribution}
\end{figure}

We present the equivalent width (EW) distribution of our HAE samples in Figure \ref{fig:EW_distribution} (left panel).  The EWs are determined by
\begin{equation}
\mathrm{EW}=\int_{\lambda_0}^{\lambda_1} \frac{f_{\text {line }}}{f_{\mathrm{cont}}} \mathrm{d} \lambda \approx \frac{F_{\text {line }}}{f_{\mathrm{cont}}} ,
\label{eq:EW_cal}
\end{equation}
where $\lambda_0$, $\lambda_1$ define the wavelength range for integration, $F_{\rm line}$ is the emission line flux and $f_{\rm cont}$ is the rest-frame continuum flux density. For \ha, we obtain $f_{\rm cont}$ by fitting the 6530--6600\AA\ stellar continuum provided by the best-fit SED using power-law models, and extrapolate to the wavelength of \ha.  The \ha\  EW (EW$_{\rm H\alpha}$) of our sample ranges from 93\AA\ to 4300 \AA, with a median value of $665\pm67$\AA.

\subsection{Search for \lya\  emission}\label{sec:MUSE_lya}
We search for \lya\ emission for all the HAEs covered by MUSE observations. Instead of the blind line  algorithm usually adopted in MUSE 3D datacubes, we now know exactly the position and redshift of our targets and can well predict where the \lya\ emission lines could emerge. For each \ha\ source, we extract $20\times20\arcsec$ sub-cubes and perform median filtering, with a window size of 101 pixels to remove continuum and nearby sources. For sources in overlap regions of multiple pointings, we combine all the individual sub-cubes weighted by their exposure time\footnote{Note that the MUSE datacubes we adopt are products of different pipelines, with significant differences in estimating uncertainties. The standard pipeline underestimates the true uncertainties, beause it doesn't consider correlated noise due to resampling. Thus additional variance calibration procedures are carried out in \cite{Urrutia2019} and \cite{Bacon2023} respectively.  To avoid the systematics induced by different variance estimates, we don't use any variance information in this work.}. We then search for local surface brightness peaks on 3-pixel width pseudo-narrowband slices and $4\times4$-pixel boxes, centered on the source position and its rest-frame 1216\AA. The box size is determined by the typical size of UV star-forming region of high-redshift galaxies shown in \cite{Kusakabe2022}. We perform the search within a radius of 1.5\arcsec\ spatially and a spectral range of 1500 km s$^{-1}$.  
This radius ensures the including of potential Ly$\alpha$ flux 
given a possible offset between the Ly$\alpha$ and H$\alpha$ due to gas distribution and radiation transfer \cite[e.g.,][]{Cai2019, Zhang2023}. 
For each selected flux peak, we extract a 1D spectrum  using the $4\times4$-pixel box, and fit a skewed Gaussian profile. We visually inspect the extracted spectra and identify 63 HAEs with \lya\ emission lines detected at $>2\sigma$. In the following analysis, we refer to the HAEs with \lya\ lines detected as `LAEs' and the remaining as `non-LAEs'; but we note that this definition fully depends on the depth of the MUSE observations and are not comparable to the literature definitions (e.g., \lya\ EW $>10\AA$). Among the 63 LAEs, 39 have been reported in \cite{Kerutt2022} or \cite{Bacon2023}.

\begin{figure*}[htbp!]
    \centering
    \includegraphics[width=\columnwidth]{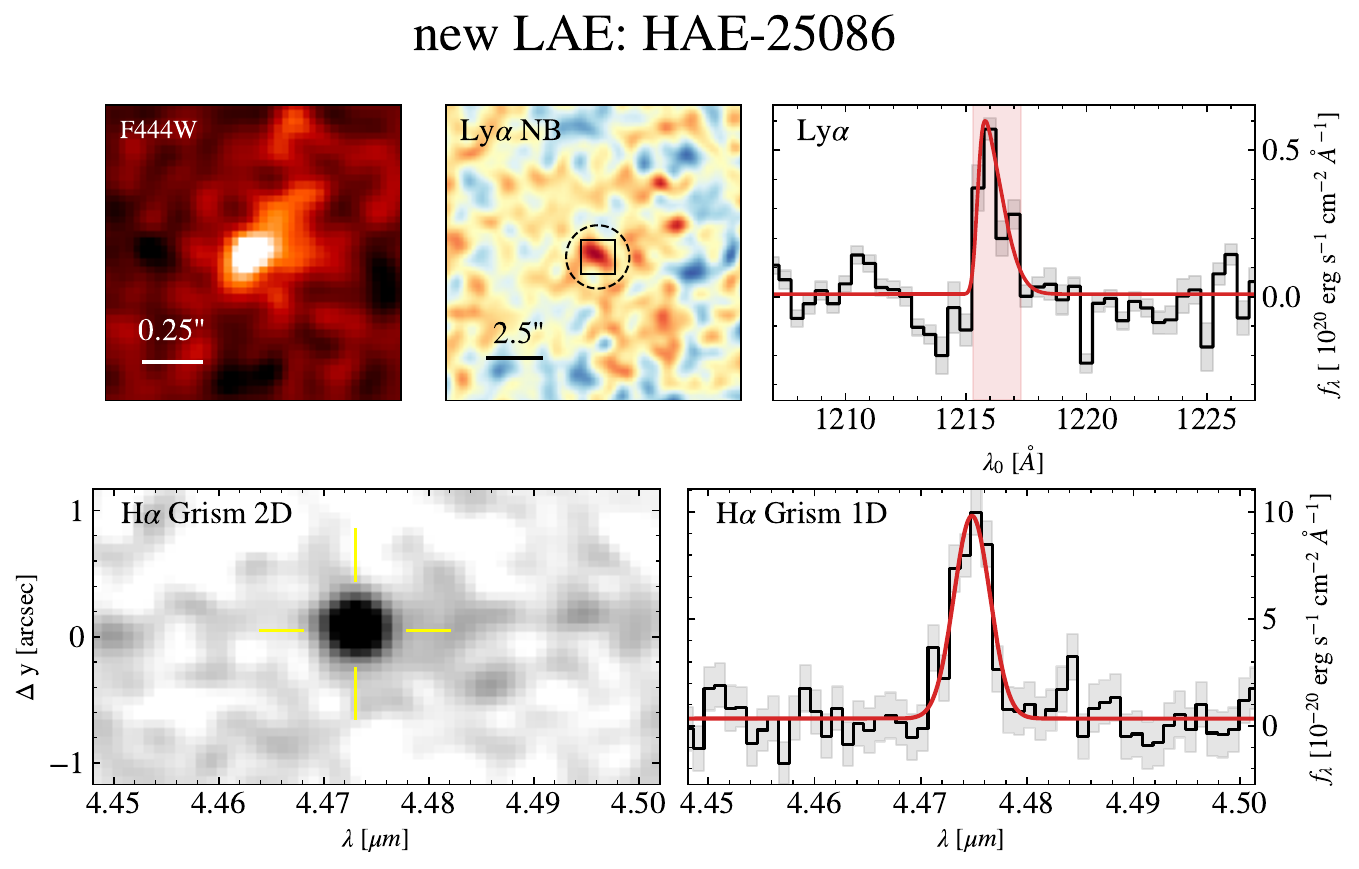}
    \includegraphics[width=\columnwidth]{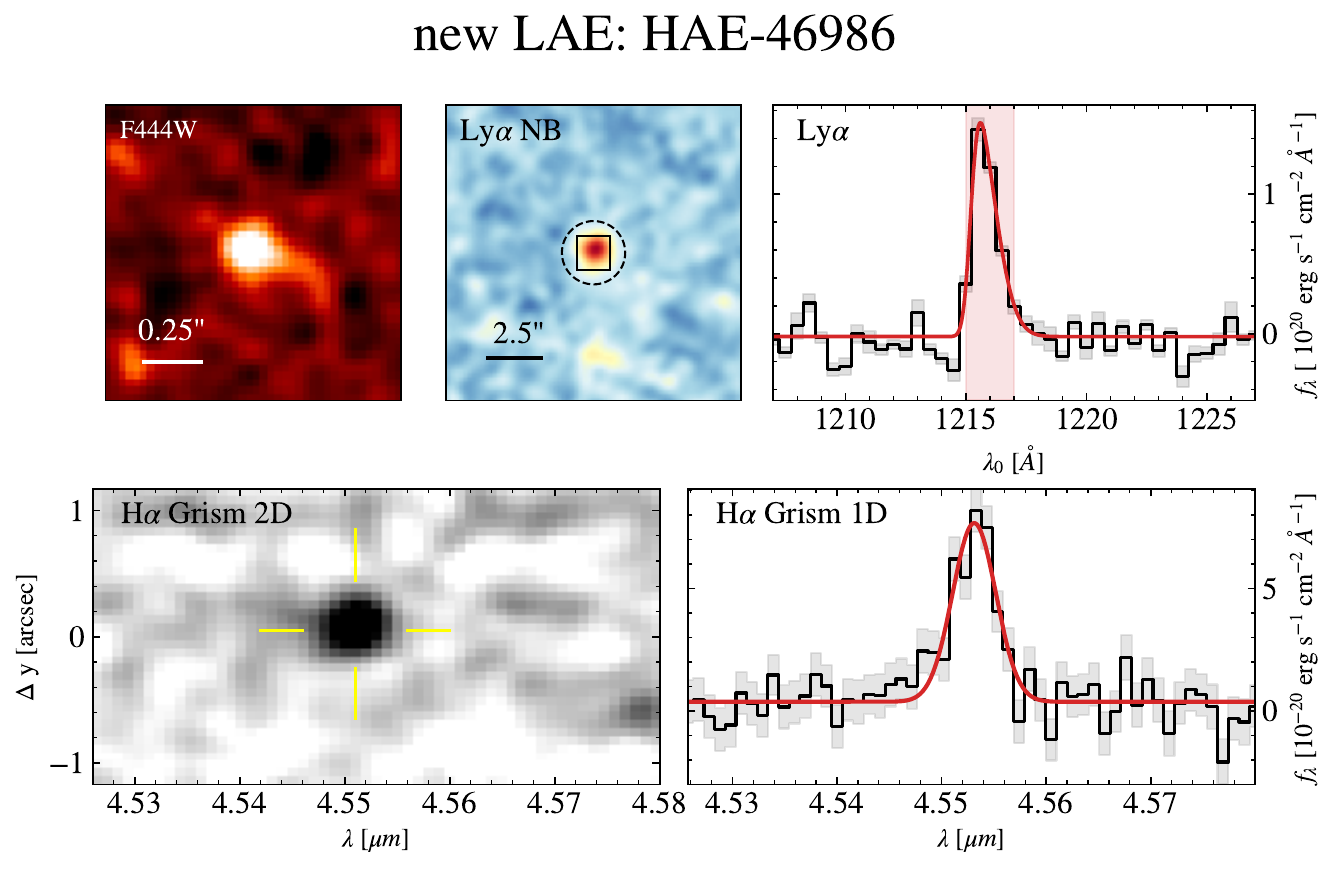}
    \includegraphics[width=\columnwidth]{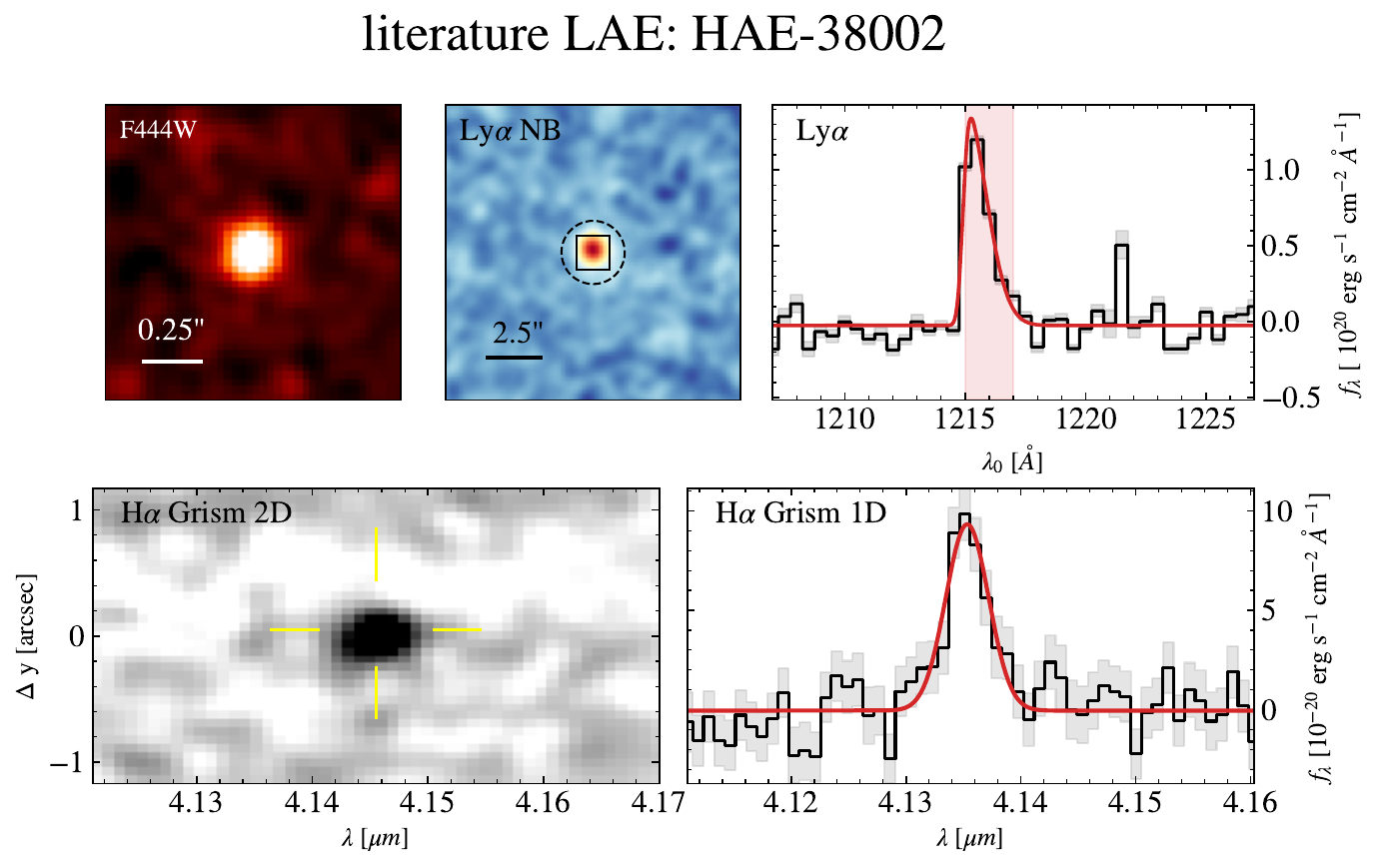}
    \includegraphics[width=\columnwidth]{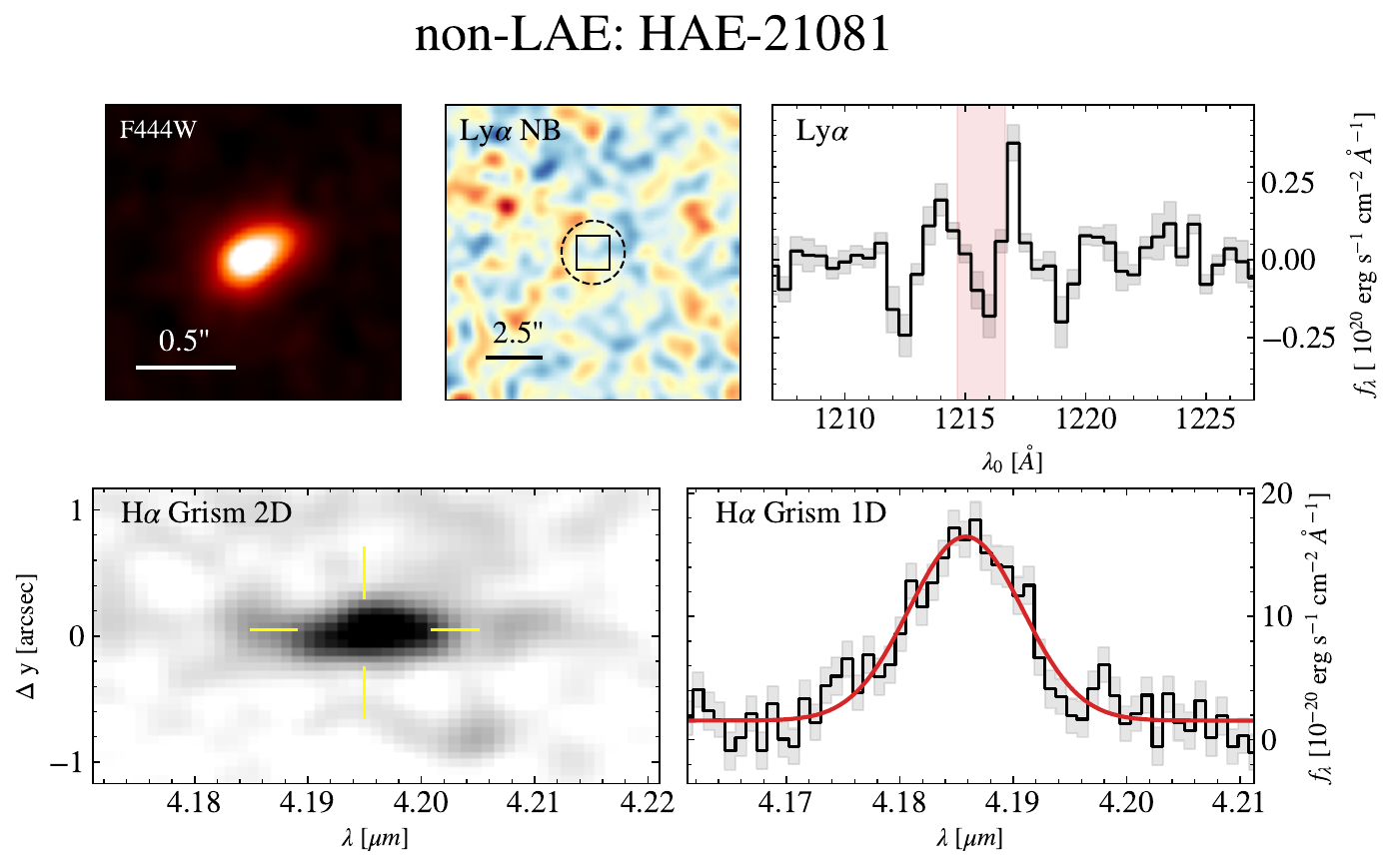}
    \caption{HAE examples with their JWST/NIRcam F444W images (top first), psedo-\lya\ narrowband images (top second), and their \lya\ (top third), \ha\ spectra (bottom). 
    The solid box on the NB image is the $4\times 4$-pixel boxy aperture used to identify \lya\ and fit its profile; the dashed circle is the $r=1.5^{\prime\prime}$ aperture for the total flux. The top third is the boxy-aperture extracted spectrum (black) and its uncertainty (grey-shaded). We show the best-fit skewed Gaussian profile in red. The red-shaded region refers to the spectral window (3-pixel width) for \lya\ NB . 
    The bottom panels of each example demonstrate the 2D and 1D grism spectrum for \ha. The yellow cross marks the position of \ha\ emission in the 2D grism spectrum and the best-fit Gaussian profile is presented in red. 
    } 
    \label{fig:LAE_HAE_example}
\end{figure*}

We further extract 1D spectra for all the selected \nHAE\  HAEs using $r=$1.5\arcsec circular apertures, and measure the uncertainties by the standard deviation of 500 spectra extracted using equally-sized apertures randomly positioned in each sub-cube. For LAEs, we fix the skewed Gaussian profile derived from the boxy-aperture extracted spectra to maximize SNRs, and adjust the amplitude to fit the circular-aperture extracted spectra and obtain the total flux.  \lya\ of two LAEs are overwhelmed by noises in their circular-aperture extracted spectra; we thus adopt the boxy-aperture measured flux for them. Comparing the flux of LAEs we measure with the values reported in \cite{Kerutt2022} and \cite{Bacon2023}, the median flux ratio is $1.04\pm0.38$ implying a good consistency. For non-LAE, we measure the integrated flux and uncertainties over  $\pm 1000$ km s$^{-1}$   around the wavelength of \lya\ within the circular apertures. All non-LAEs do not show emission line peaks around the \lya\ wavelength and have integrated \lya\ fluxes  with SNRs$<3$. Figure \ref{fig:LAE_HAE_example} shows six examples of LAEs and non-LAEs among our HAE sample. 
We show the distribution of \lya\ EW ($\rm EW_{Ly\alpha}$) of LAEs and non-LAEs in the right panel of Figure \ref{fig:EW_distribution}. $\rm EW_{Ly\alpha}$ are measured following Equation \ref{eq:EW_cal}, with $f_{\rm cont}$ obtained by fitting the 1250--1350\AA\ stellar continuum.  Our detected LAEs have rest-frame EW ($\rm EW_{Ly\alpha}$) spanning from 19\AA\ to 462\AA, with a median of 76\AA. For non-LAE, $\rm EW_{Ly\alpha}$ derived from the forced measured \lya\ within $\pm$1000 km $^{-1}$ spans from -175\AA\ to 441\AA, with a median value of 11\AA. 
Their $3\sigma$ upper limits range  
from 8\AA\ to 1123 \AA\ with a median value of 91 \AA.


\subsection{ Measurement of \lya\ escape fraction }\label{sec:measure_fesclya}
In this section we describe our method to measure \fesclya\ in detail.  We outline the measurement for individual HAEs in \S\ref{sec:fesc individual} and the stack procedure in \S\ref{sec:lya stack}, which is aimed at measuring the overall \fesclya\ of a specific population.

\subsubsection{ \fesclya\ of individual HAEs}\label{sec:fesc individual}

Under the assumption of Case B recombination for $T=10^4$K and $n_e\approx350$ cm$^{-3}$ gas,  the \lya\ escape fraction could be expressed as \citep{Osterbrock1989}
\begin{equation}\label{eq:fesclya}
    f_{\rm esc,Ly\alpha} = \frac{F_{\rm Ly\alpha, obs} }{8.7 {F}_{\rm H\alpha,int}}
\end{equation}
where  ${F}_{\rm Ly\alpha,obs}$ is the observed \lya\ flux and  ${F}_{\rm H\alpha,int}$ is the intrinsic \ha\ flux where the dust attenuation 
effect has been corrected. For LAEs, \fesclya\ could be directly calculated following Equation \ref{eq:fesclya} with the detected \lya\ emission flux as described in \S\ref{sec:MUSE_lya}; for non-LAEs, we use the integrated flux over $\pm1000$ km s$^{-1}$ within the $r=1.5\arcsec$ circular aperture.  We correct the dust attenuation with the \citetalias{Calzetti2000} and SMC laws and the optical depth in the V band derived from \texttt{Beagle} respectively. The differential dust attenuation factor between stars and nebular regions is set as $f_{\rm nebular}=1$, 
as suggested by a number of previous studies that $f_{\rm nebular}$ approaches unity at $z>2$ with \ha\ EW $>100\AA$ \cite[e.g.,][]{Faisst2019}.  The distribution of \fesclya\ of individual HAEs will be discussed in \S\ref{sec:distribution_fesclya}


\subsubsection{\fesclya\ by VLT/MUSE stack}\label{sec:lya stack}

To calculate the typical \fesclya\ of a specific population comprising HAEs with both \lya\ detected and not detected, we stack the MUSE datacubes and measure the stacked \lya. We shift the median-filtered sub-cubes of each HAE to its rest frame and then resample the sub-cubes into a fixed grid along the spectral axis. The rest-frame sub-cubes are then normalized based on their intrinsic \ha\ fluxes, with each voxel in the cube divided by the intrinsic \ha\ flux. We obtain the median-stacked cube by taking the median value of each voxel element to avoid the effect of outliers and asymmetric distribution of \fesclya. The uncertainty of each voxel is estimated by bootstrapping sampling a thousand times.  

We extract 1D \lya\ spectra using a $r=2.5\arcsec$ circular aperture on the median-stacked cubes. This aperture 
is conservative, 
 larger than what we use for individual HAEs, as we randomly align the subcubes during the stack without considering the asymmetry of \lya\, and thus the stacked \lya\ could be more extended than individuals. On the \lya\ psedo-narrowband image of the stack of all LAEs integrated over $\pm 1000\ {\rm km\ s^{-1}}$, an $r=2.5\arcsec$ aperture includes $95\pm12\%$ of the total flux. We fit skewed Gaussian profiles on the extracted 1D spectra to estimate the normalized \lya\ flux. 

\begin{figure}
    \centering
    \includegraphics[width=\columnwidth]{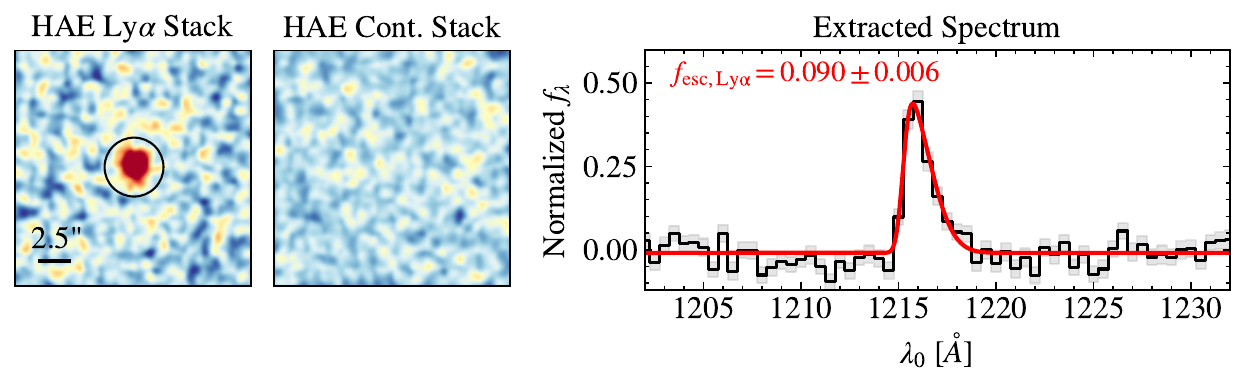}
    \includegraphics[width=\columnwidth]{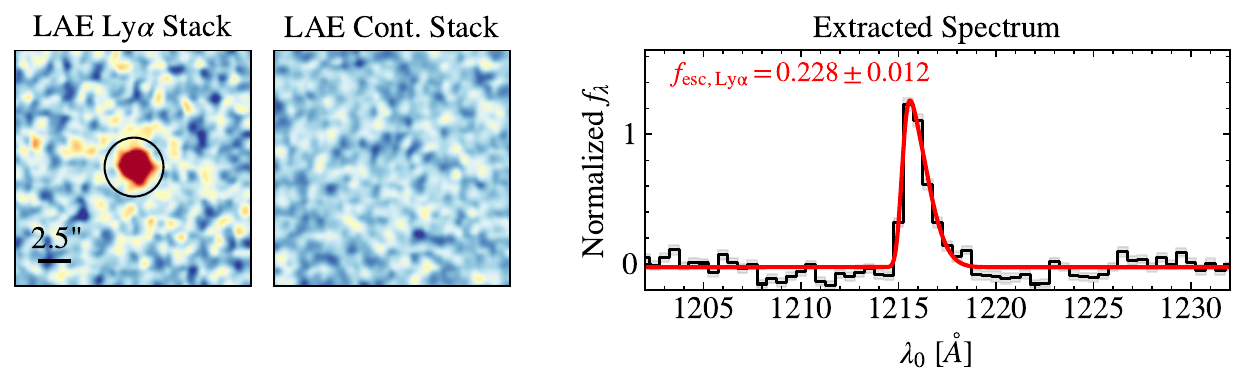}
    \includegraphics[width=\columnwidth]{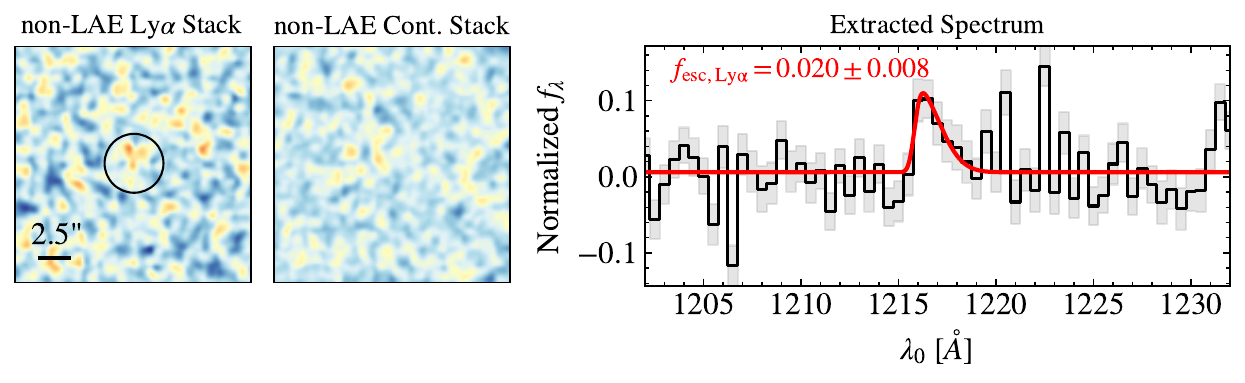}
    \caption{\lya\ stacks of MUSE datacubes for all HAEs in our sample,  as well as the LAE and non-LAE subsamples. All datacubes are normalized by the intrinsic \ha\ flux following the \citetalias{Calzetti2000} dust attenuation law. The psedo-\lya\ narrowband (NB) images in the left cover 1213--1219\AA, i.e. slices of $\pm 3\AA$ centered on 1216\AA; the continuum (Cont. in short) NB images are calculated from 1230--1240\AA\ slices and scale to the wavelength width of \lya\ slices. The extracted spectra based on $r$=2.5\arcsec circular apertures (black solid circle in \lya\ NB images) are shown, with the best-fit skewed Gaussian profiles in red. 
    }
    \label{fig:lya_stack}
\end{figure}

In Figure \ref{fig:lya_stack}, we show \lya\ stacks of all HAEs, LAEs, and non-LAEs.  The overall \fesclya\ for these HAEs is \fesclyaCal (\fesclyaSMC) under the assumption of the \citetalias{Calzetti2000} (SMC) dust attenuation law. For LAEs, \fesclya\ is $0.228\pm0.012$ ($0.275\pm0.013$); for non-LAE, \fesclya\ is $0.020\pm0.008$ ($0.030\pm0.011$). The \fesclya\ of LAEs are over twice the average \fesclya\ of all HAEs, strongly indicating that previous studies based on \lya-selected galaxies overestimate the average \fesclya\ of the whole galaxy populations.

%% file: 04result.tex
\section{Results}\label{sec:result}
We present the results of our \fesclya\ measurements as follows: we describe the distribution of \fesclya\ in \S\ref{sec:distribution_fesclya}; the dependence of \fesclya\ on various galaxy properties is discussed in \S\ref{sec:fesclya_galaxy_properties}; in \S\ref{sec:redshift} we show the redshift evolution of \fesclya.

\subsection{The distribution of \fesclya}\label{sec:distribution_fesclya}

\begin{figure}
    \centering
    \includegraphics[width=\columnwidth]{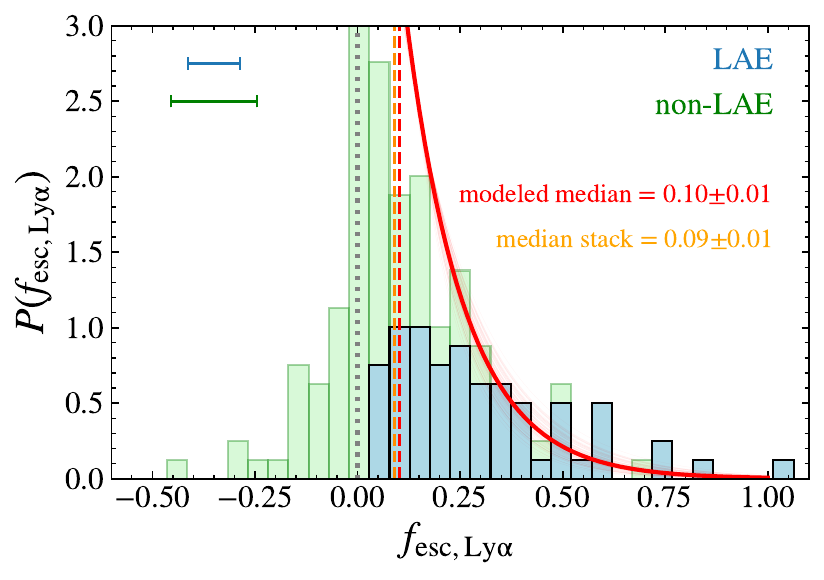}
    \caption{The distribution of \fesclya\ for LAEs and nonLAEs, assuming the \citetalias{Calzetti2000} dust attenuation law.  For non-LAEs, we show the forced measured \fesclya\ based on the integrated flux over $\pm 1000{\rm km/s}$ around the wavelength of \lya. The mean errorbar of \fesclya\ of LAEs and non-LAEs are marked as the blue and green horizontal lines. The thick red curve is the best-fit exponential model for the \fesclya\ distribution; the thin red curves represent 100 exponential models with $f_*$ (Equation \ref{eq:P_fesc}) randomly drawn from the \textsc{MCMC} samples. \fesclya$=0$ is labeled as the gray dotted line. The red dashed line denotes the median \fesclya\ predicted by the best-fit exponential model, and the orange dashed line is the \fesclya\ measured from the stack of all HAEs. }
    \label{fig:fesc_distribution}
\end{figure}

We measure the \fesclya\ of individual HAE following the method in \S\ref{sec:fesc individual}. The distributions of \fesclya\ of LAEs and nonLAEs are shown in Figure \ref{fig:fesc_distribution}. For our detected LAEs, \fesclya\ ranges from  0.04 to 1. For non-LAEs, there are negative \fesclya\  measurements with large uncertainties; the negative values are all consistent with zero within $3\sigma$. The mean and median values of the individual \fesclya\ are $0.13\pm 0.01$ and $0.08\pm 0.01$ ($0.15\pm0.01$ and $0.11\pm0.01$) under the assumption of \citetalias{Calzetti2000} (SMC) dust attenuation law. We model the distribution of \fesclya\ using an exponential model, as commonly adopted to depict the distributions of \lya\ EWs of \lya\ emitting galaxies \citep[e.g.,][]{Dijkstra2010, Dijkstra2012, Kerutt2022}. In fact \fesclya\ is proportional to the ratio of \lya\ and the \ha\ fluxes, as shown in Equation \ref{eq:fesclya}; meanwhile \lya\ EWs are the ratios between \lya\ fluxes and the underlying UV continua. \ha\ fluxes and UV continua both trace star formation so that \fesclya\ and \lya\ EWs are correlated \citep[e.g.,][]{Yang2016,Sobral2019}.  The distribution of \fesclya\ is expressed as:

 \begin{equation}
\label{eq:P_fesc}
P(f_{\rm esc,Ly\alpha}|f_*) =\left\{
\begin{aligned}
& 0, ~ f_{\rm esc,Ly\alpha} < 0 {~\rm or ~} f_{\rm esc,Ly\alpha} > 1,~\\
& \propto  e^{-f_{\rm esc,Ly\alpha} / f_*}, {~\rm otherwise}.
\end{aligned}
\right.
\end{equation}
where $\int_{0}^{1} P(f_{\rm esc,Ly\alpha}|f_*)df_{\rm esc,Ly\alpha}=1$, $f_*$ is the characteristic (mean) escape fraction. 

For both LAEs and non-LAEs with measured escape fraction $\hat{f}_{\rm esc,Ly\alpha}$ and uncertainty $\Delta \hat{f}_{\rm esc,Ly\alpha}$, we assume the uncertainty is Gaussian and the probability for an \lya\ escape fraction being $\hat{f}_{\rm esc,Ly\alpha}$ should be 
\begin{equation}
\begin{split}
     & \mathcal{P}(\hat{f}_{\rm esc,Ly\alpha},  \Delta \hat{f}_{\rm esc,Ly\alpha}|f_*) \\
     &= \int_{0}^{1} P(x|f_*) P_{\mathcal{N}}\left(x,\hat{f}_{\rm esc, Ly\alpha}, \Delta\hat{f}_{\rm esc,Ly\alpha}\right) dx
\end{split}
     \label{eq:p_fesc_LAE}
\end{equation}
where $P_{\mathcal{N}}\left(x,\hat{f}_{\rm esc, Ly\alpha}, \Delta\hat{f}_{\rm esc,Ly\alpha}\right)$ is the probability of the escape fraction being measured as $\hat{f}_{\rm esc, Ly\alpha}$ if the true value is $x$ with the uncertainty $\Delta\hat{f}_{\rm esc,Ly\alpha}$:
\begin{align*}
    P_{\mathcal{N}}\left(x,\hat{f}_{\rm esc, Ly\alpha}, \Delta\hat{f}_{\rm esc,Ly\alpha}\right) &= \frac{1}{(2 \pi)^{1 / 2}  \Delta \hat{f}_{\rm esc,Ly\alpha}} \\
    & \times \exp \left[-\frac{1}{2}\left(\frac{x-\hat{f}_{\rm esc,Ly\alpha}}{\Delta \hat{f}_{\rm esc,Ly\alpha}}\right)^2\right].
\end{align*}
The overall likelihood for our observed \fesclya\ distribution should be a product of the \fesclya\ likelihoods of all galaxies. With a flat prior of $f_*$ ranging from zero to one,  we perform Bayesian inference using the Markov chain Monte Carlo 
 method (\textsc{MCMC\footnote{ We use \textsc{emcee} \citep{emcee} to perform MCMC: \url{https://emcee.readthedocs.io/en/stable/}}}) to model the observed \fesclya\ distribution. It yields the best-fit $f_*=0.15\pm0.02$ ($0.17^{+0.03}_{-0.02}$) for \citetalias{Calzetti2000} (SMC) dust law, in agreement with the mean value of \fesclya\ ($0.13\pm0.01$ for \citetalias{Calzetti2000} and $0.15\pm0.01$ for SMC). The median \fesclya\ value predicted by the best-fit exponential distribution is  $f_* \ln 2$, i.e. $0.10\pm0.01$ ($0.12^{+0.02}_{-0.01}$) for \citetalias{Calzetti2000} (SMC) dust law, in line with the median stack of all HAEs (\fesclyaCal\ for \citetalias{Calzetti2000}; \fesclyaSMC\ for SMC).

\subsection{The dependence of \fesclya\ on galaxy properties}\label{sec:fesclya_galaxy_properties}

\begin{figure*}
    \centering
    \includegraphics[width=\textwidth]{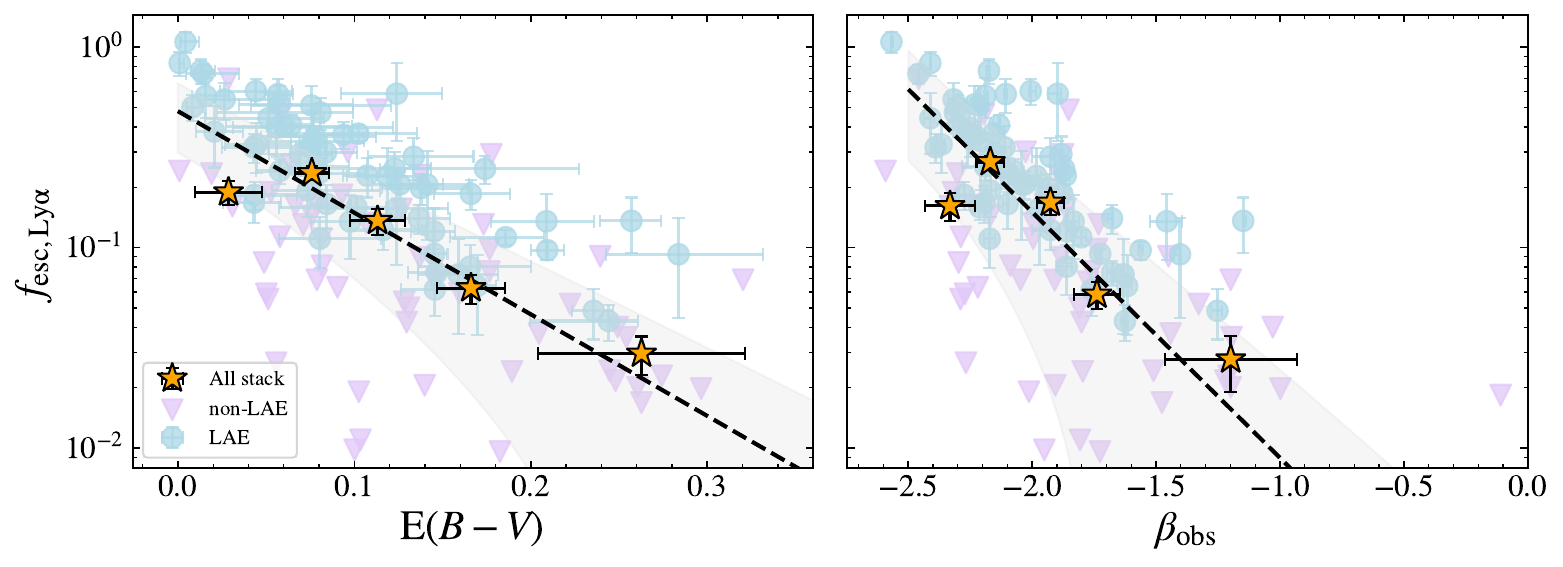}
    \caption{\fesclya\ varying with the observed UV slope \betaobs\ and \EBV\ , under the assumption of \citetalias{Calzetti2000} dust attenuation law. The light blue circles denote LAE individuals and the light purple triangles are non-LAEs with forced \fesclya\ measured from the integrated \lya\ flux within $\pm1000$ km $^{-1}$. For illustration purposes  we do not show the uncertainties of \fesclya\ of non-LAEs. The orange stars are the overall \fesclya\ of all HAEs derived from the \lya\ stack results. The dashed lines represent the best-fit relations between \fesclya\ and \EBV\  or \betaobs\ based on the stack results as described in Equation \ref{Eq:fescLyA_EBV}. We show their $1\sigma$ uncertainties as shaded regions.}
    \label{fig:fesclya_dust}
\end{figure*}

\begin{figure*}
    \centering
    \includegraphics[width=\textwidth]{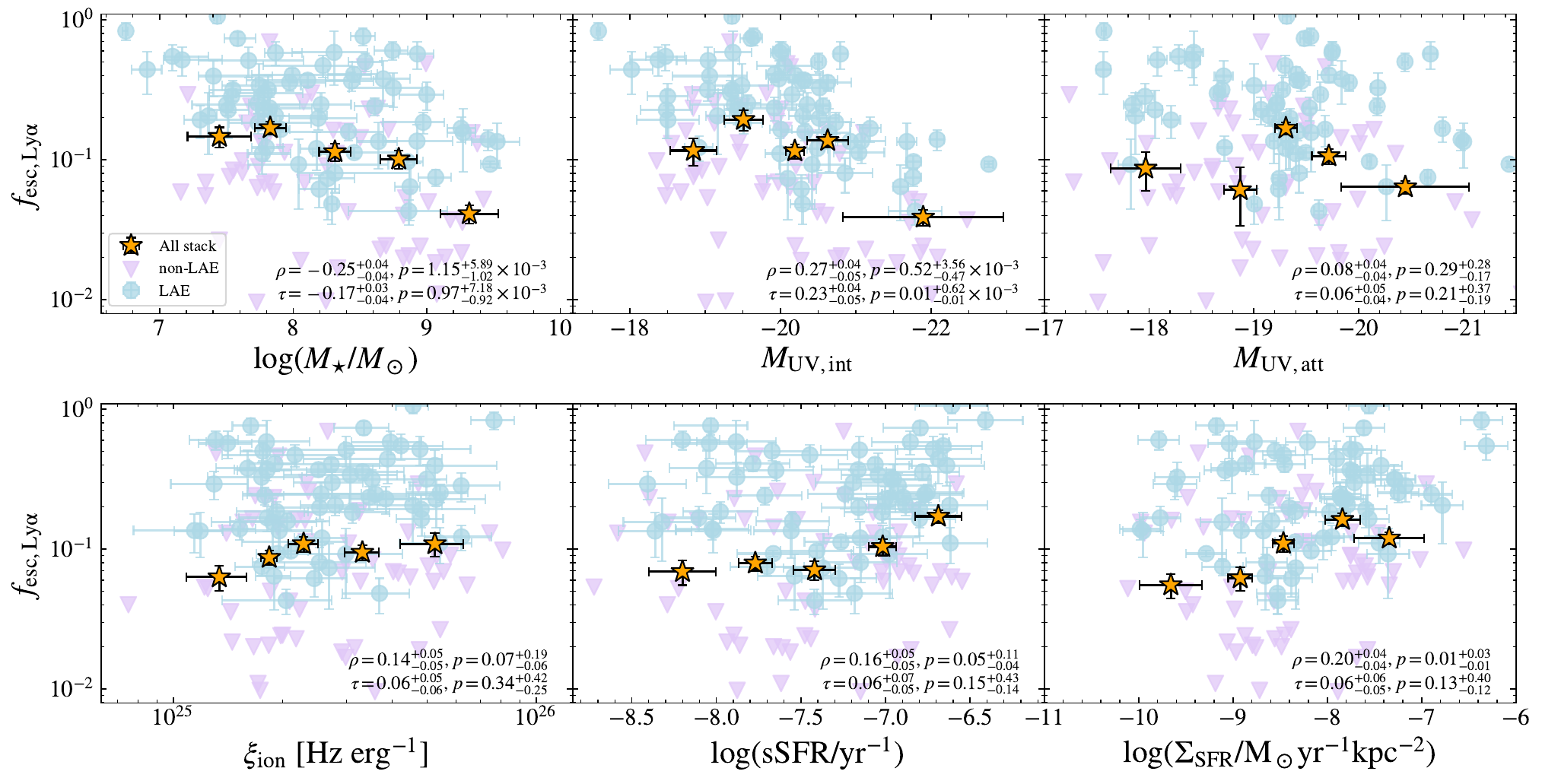}
    \caption{\fesclya\  varying with different physical parameters of HAEs assuming the \citetalias{Calzetti2000} dust attenuation law. The light blue dots denote the \fesclya\ measured for individual LAE; the light purple triangles are the forced measured \fesclya\ for non-LAEs. The uncertainties of \fesclya\ of non-LAEs are not shown for illustration purposes. The orange stars represent the stack \fesclya. We label the correlation coefficients of the Spearman's correlation analysis ($\rho$) and Kendall’s correlation analysis ($\tau$) for each property, as well as the corresponding $p$-values.  The Spearman's correlation analysis is based on the measured values and uncertainties of all individual LAEs and non-LAEs, while in the Kendall’s tests we adopt the $3\sigma$ upper limits of \fesclya\ for non-LAEs.}
    \label{fig:fesclya_parameter}
\end{figure*}

\lya\ photon transportation can be modulated by the dust and neutral gas content in galaxies, as well as the geometry and kinematics of ISM \citep{Atek2009, Smith2019, Maji2022, Tang2024}.
 We focus on nine physical properties that are most commonly studied: stellar mass (\Ms), ionizing efficiency (\xiionzero), redshift ($z$), the attenuated (observed) and the intrinsic absolute UV magnitudes(\MUVatt), and \MUVint), specific star formation rate (sSFR), sSFR surface density (\SDsSFR), the observed UV slope (\betaobs) and the dust attenuation (\EBV). The measurements of these physical parameters are described in \S\ref{sec:SED}. To investigate how \fesclya\ correlates with the properties of HAEs, we split the HAE sample into five equal-sized bins based on each physical parameter, with 32 or 33 HAEs in each bin. We then estimate the overall \fesclya\ in each bin by stacking the \lya.  The dependence of \fesclya\ on the nine physical parameters is presented in Figure \ref{fig:fesclya_dust} and Figure \ref{fig:fesclya_parameter}. To test whether there are monotonic relations between \fesclya\ and these physical parameters,
 we perform the non-parametric Spearman's correlation analysis\footnote{\url{https://github.com/privong/pymccorrelation}} using a Monte Carlo approach following \cite{Curran2014}.
 We include the forced measured \fesclya\ of non-LAEs and the corresponding uncertainties in the analysis for unbiased statistics. We also test the correlations with Kendall rank correlation analysis \citep{Isobe1986}, where the $3\sigma$ upper limits of non-LAEs' \fesclya, instead of their forced measurements, are used as censored data in the analysis. As shown in Figure \ref{fig:fesclya_dust}, \fesclya\ shows a strong dependence on \betaobs\ and \EBV\ , with a Spearman's correlation coefficient ($\rho$) of $-0.37\pm0.04$ ($p$-value \footnote{$p$ is the p-value indicating the probability of obtaining the current result if the correlation coefficient were zero (no correlation). If $p$ is lower than 0.05, the correlation coefficient is statistically significant.} $= 0.17^{+2.21}_{-0.16}\times 10^{-5}$) and $-0.34\pm0.05$ ($p=0.08^{+1.11}_{-0.07}\times 10^{-4}$). The Kendall correlation analysis shows correlation coefficients ($\tau$) of $-0.34\pm0.04$ with $p\approx 10^{-10}$ and $-0.37\pm0.03$ with $p\approx 4\times 10^{-12}$ for \fesclya$-$\betaobs\ and \fesclya$-$\EBV\  respectively. We quote the median values of $\rho$, $\tau$, and $p$ and their uncertainties correspond to the 16 and 84 percentiles.  For \Ms, \MUVint, \MUVatt, \xiion, sSFR and \SDsSFR, we illustrate their correlations with respect to \fesclya\ Figure \ref{fig:fesclya_parameter}, and label the correlation coefficients of Spearman's and Kendall's tests as well as their corresponding  $p$-values.  We find that \fesclya\ also weakly correlates with \Ms\ and \MUVint, with $p$ of both the two correlation tests much smaller than 0.05. For \MUVatt, the correlation coefficients, $\rho$ and $\tau$, are both close to zero with large $p$-values, so \fesclya\ is almost not related to \MUVatt. In the cases of \xiion, sSFR, and \SDsSFR,  the $p$-values of Spearman's correlation coefficients can be comparable to or exceed a lot 0.05 within 1$\sigma$ ranges. The large uncertainties in $p$ are a result of the large uncertainties of the forced measured \fesclya\ of non-LAEs. On the other hand, their Kendall's correlation analysis reveals $\tau$ close to zero with large $p$-values, indicating no correlations these quantities.
 
 
 
In conclusion, our results demonstrate that \fesclya\ has a strong dependence on \betaobs\ and \EBV\ . \fesclya\ is also monotonically related to \Ms\ and \MUVint,  but no correlations can be confirmed between \fesclya\ and \MUVatt, \xiion, $\log$sSFR and \SDsSFR.

In fact, \EBV\  and \betaobs\ have long been thought to be the most crucial factors to trace \fesclya\ from low redshifts \citep[e.g.,][]{Atek2009,Yang2016} to cosmic noon \citep[e.g.,][]{Hayes2010, Sobral2019}.  We present their relationship with respect to \fesclya, for the first time, at redshift $z>5$. As shown in Figure \ref{fig:fesclya_dust}, \fesclya\ appears remarkably anti-correlated with \EBV\  and \betaobs. We fit the relation using the orthogonal distance regression method\footnote{\url{https://docs.scipy.org/doc/scipy/reference/odr.html}} with the stack \fesclya\ in Figure \ref{fig:fesclya_dust}, assuming

\begin{equation}
\label{Eq:fescLyA_EBV}
f_{\rm esc,Ly\alpha} =\left\{
\begin{aligned}
& f_{\rm Ly\alpha,0} \cdot 10^{- 0.4 k_{\rm Ly\alpha} {{\rm E}(B-V) }},  \\
& f_{\rm Ly\alpha,0}^{\prime }   \cdot  10^{-\alpha_{\rm Ly\alpha}   {(\beta_{\rm obs} + 2.5)}} ,
\end{aligned}
\right.
\end{equation}
For \citetalias{Calzetti2000} (SMC) law, we obtain $f_{\rm Ly\alpha,0}=0.48\pm0.18\ (0.46\pm0.19)$,  $k_{\rm Ly\alpha}=12.68\pm 4.07\ (15.21\pm5.73)$. For the \fesclya$-$\betaobs\ relation, we obtain the best-fit $f_{\rm Ly\alpha,0}^{\prime }=0.62\pm0.34\ (0.54\pm0.19)$ and $\alpha_{\rm Ly\alpha} = 1.23\pm0.48\ (0.71\pm0.21)$.  This indicates that dust-free galaxies at $z=5-6$, on average, exhibits \fesclya\ of $0.48\pm0.18\ (0.46\pm0.19)$;  galaxies with blue UV slopes $\beta\sim -2.5$ 
have \fesclya\ around $0.62\pm0.34\ (0.54\pm0.19)$.
 
\xj{Among the 165 HAEs, we note that two LAEs (HAE-60916 and HAE-87611) exhibit double-peaked \lya\ profiles (Figure \ref{fig:double_peak_LAE}, see Appendix \ref{sec:double_peak_LAE} for more details). They have blue UV slopes, \betaobs$\approx-2.107\pm0.007$ and $-1.871\pm0.023$, and high \lya\ escape fractions, \fesclya$\approx 0.58\pm0.11$ and $0.25\pm0.04$. The predicted $f_{\rm esc, Ly\alpha}$  based on their \betaobs\ using Equation \ref{Eq:fescLyA_EBV} is $0.20\pm0.14$ and $0.10\pm0.09$ respectively. The observed  \fesclya\ of the two  double-peaked LAEs are $\sim 2\sigma$ higher than the predicted values. The presence of blue \lya\ peaks is  likely to signal a boosted escape of \lya\ photons compared to normal star-forming galaxies, consistent with previous findings on double-peaked LAEs at the cosmic noon \citep{Marques2022, Furtak2022}. Studies on statistical samples with high SNR \lya\ profiles are expected in the future using deep spectroscopy. }

\subsection{Redshift evolution of \fesclya}\label{sec:redshift}

\begin{figure}
    \centering
    \includegraphics[width=\columnwidth]
    {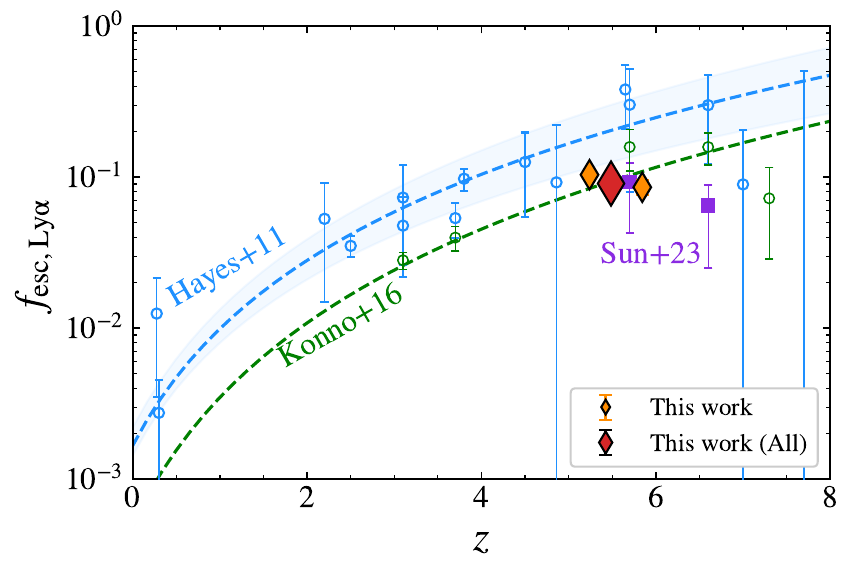}
    \caption{The redshift evolution of global \fesclya\ from $z=0$ to $z=8$. Our stacked measurement of \fesclya\ for all HAE samples is shown as a red diamond based on the assumption of \citetalias{Calzetti2000} dust law. The orange and white diamonds are \fesclya\ measured by stacking the HAE sample in two redshift bins ($z=5.24,\ 5.84$). 
    \fesclya\ measurements compiled by \cite{Hayes2011} based on \lya\ and UV ($z>2.2$) or \ha\ luminosity densities ($z<2.2$) are shown in blue circles, and the best-fit relation for their compilation is shown as the blue dashed line. The green circles denote the measurements in \cite{Konno2016}, also using \lya\ and UV LFs, and their best-fit relation is shown as the green dashed line. The purple squares are the \fesclya\ derived from \lya\ and \ha\ luminosity densities at $z>5$ in \cite{Sun2022b} with HAEs observed by NIRCam/WFSS.}
    \label{fig:fescLyA_z}
\end{figure}

We study the evolution of \fesclya\ as a function of redshift as shown in \ref{fig:fescLyA_z}. We split the HAE sample into two redshift bins, $z=5.24$ ($z=4,9-5.5$) and $z=5.84$ ($z=5.5-6.2$), and then measured the stacked \lya\ respectively. The observed \fesclya\ is $0.104\pm 0.007$ at $z=5.24$ and becomes $0.086\pm0.010$ at $z=5.84$.  We compare our measurements with the \fesclya\ measurements in the literature. The observed \fesclya\ values are only about $56\%$ and $37\%$ of the estimates in \cite{Hayes2011} at $z=5.4$ and $z=5.8$ respectively. 
On the other hand, our measurements are generally consistent with the best-fit function in \cite{Konno2016} (green dashed line in Figure \ref{fig:fescLyA_z}), 
although they are still lower than their measurement at this redshift range (green circle in Figure \ref{fig:fescLyA_z}).  We note that, because \ha\ was not accessible at $z=5-6$ until the launch of JWST, all previous measurements rely on UV luminosity function (LFs) to estimate the star formation rate density and \ha\ luminosity density at these redshifts, while our results present a direct measurement for the first time. The offset between \cite{Hayes2011} and \cite{Konno2016} is due to the differences of the \lya\ and UV luminosity limits for deriving the  \lya\ and UV luminosity densities from their luminosity functions: \cite{Hayes2011} integrated \lya\ LFs from  $L_{\rm Ly\alpha}=10^{41.3}$ erg s$^{-1}$  to $L_{\rm Ly\alpha}=10^{42.9}$  erg s$^{-1}$ and UV LFs down to -17.6, while \cite{Konno2016} integrated \lya\ LFs from $L_{\rm Ly\alpha}=10^{41.7}$ erg s$^{-1}$ to $L_{\rm Ly\alpha}=10^{44.4}$ erg s$^{-1}$ and UV LFs down to -17.0 mag. In addition to this, the measurements of faint-end LFs and the conversion between UV and \ha\ SFR could 
lead to large systematic uncertainties in these UV-based \fesclya\ estimates \citep{Konno2016, Sun2022b}. The observed \fesclya\ are in good agreement with \cite{Sun2022b}, where \fesclya\ is derived from \ha\ luminosity densities of HAEs also captured by JWST/NIRCam WFSS. Our results, combined with those in \cite{Sun2022b}, suggest a tentative declining trend for the observed  \fesclya\ at $z\gtrsim5$. The average \EBV\  of our HAE sample is about 0.14 at $z=5.24$ and 0.09 at $z=5.84$, implying an intrinsically higher \fesclya\ at the higher redshift due to their less dust content. The decline of \fesclya\ might be a result of the increasing neutral fraction of IGM as $z$ increases. To completely understand the redshift evolution of \fesclya, direct observations on \lya\ and \ha\ emission lines across wide redshift spans are required in the future.


%% file: 05discussion.tex
\section{Discussion}\label{sec:discussion}

In this section, we discuss the physical origins of the correlations as outlined in \S\ref{sec:result}, connect the escape of \lya\ 
and LyC photons in \S\ref{sec:lya and lyc},  and further qualitatively constrain ionizing photon budget for the reionziation in \S\ref{sec:reionization}.

\subsection{Implications for \lya\ and LyC escape}\label{sec:lya and lyc}

\begin{figure}
    \centering
    \includegraphics[width=\columnwidth]{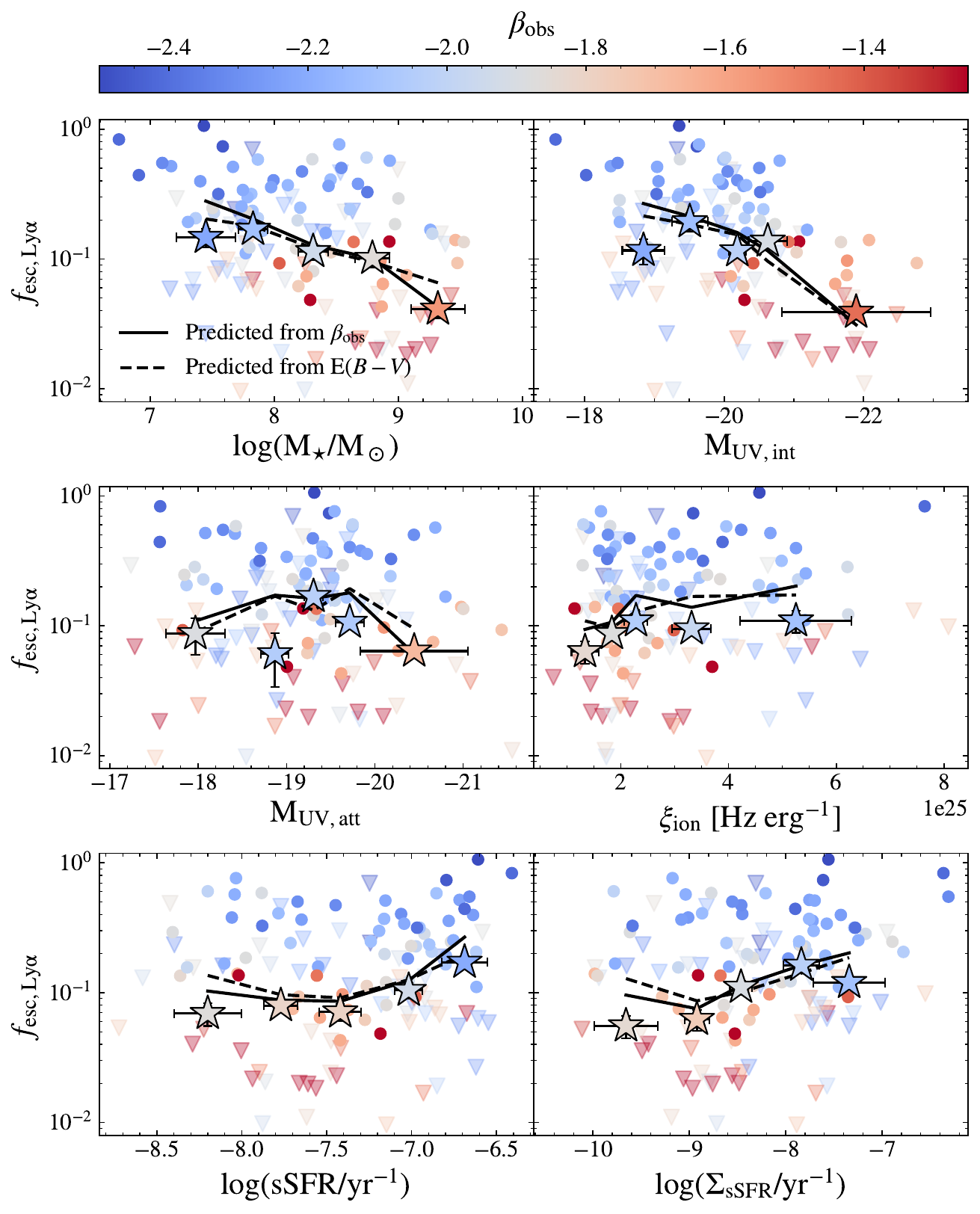}
    \caption{The coupling of \betaobs\ and \EBV\  on the dependence of \fesclya\ on \Ms, \MUVatt, \MUVatt, \xiion, sSFR and \SDsSFR. We show  the \fesclya\ of individual LAEs as circles and forced measured \fesclya\ of non-LAEs as triangles, color-coded by their \betaobs. The stack \fesclya\ are marked as stars color-coded by the median \betaobs\ in each stack bin. The solid and dashed lines are predicted \fesclya\ from Equation \ref{Eq:fescLyA_EBV} based on the median \betaobs\ and \EBV\   in each stack bin, respectively.}
    \label{fig:beta_EBV_coupling}
\end{figure}

As illustrated in \S\ref{sec:fesclya_galaxy_properties}, \fesclya\ correlates with \Ms\ and \MUVint, albeit to a lesser extent compared to the strong dependence of \fesclya\ on \betaobs\ and \EBV.  However, \Ms\ and \MUVint\ may be highly coupled with \betaobs\ and \EBV. \MUVint\ are the products of observed \MUVatt\ corrected by the effect of dust attenuation, so it is highly dependent on the dust attenuation parameter adopted. We explore the coupling of \fesclya$-$\Ms\ and \fesclya$-$\MUVint\ relations with \betaobs\ and \EBV\  in Figure \ref{fig:beta_EBV_coupling}.  Comparing the expected \fesclya\ by Equation \ref{Eq:fescLyA_EBV} based on the median \betaobs\ and \EBV\  in each \Ms\ and \MUVint\ bin, we find that the variation of \fesclya\ across \Ms\ and \MUVint\ are consistent with the prediction of Equation \ref{Eq:fescLyA_EBV}. It implies that both the \fesclya$-$\Ms\ and \fesclya$-$\MUVint\ correlations are modulated by the underlying coupling among \betaobs, \EBV\  and \Ms, \MUVint. Likewise, we examine the variations of \fesclya\ versus \MUVatt, sSFR, \SDsSFR\ and \xiion, and find that the trends are determined by the variations of median \betaobs\ and \EBV\  in each stack bin. As illustrated in Figure \ref{fig:beta_EBV_coupling}, we can reproduce the trend in \fesclya\ versus these properties by applying the median values of \betaobs\ and \EBV\ to Equation \ref{Eq:fescLyA_EBV}. To disentangle their impacts on \fesclya,  it is necessary to track the evolution of \fesclya\ within a fixed \Ms\ or \MUVint\ bin with varying \betaobs\ and \EBV\ . This requires a large sample with direct and accurate \lya\ and \ha\ detection, which cannot be achieved with our current sample. 

\begin{figure}
    \centering
    \includegraphics[width=\columnwidth]{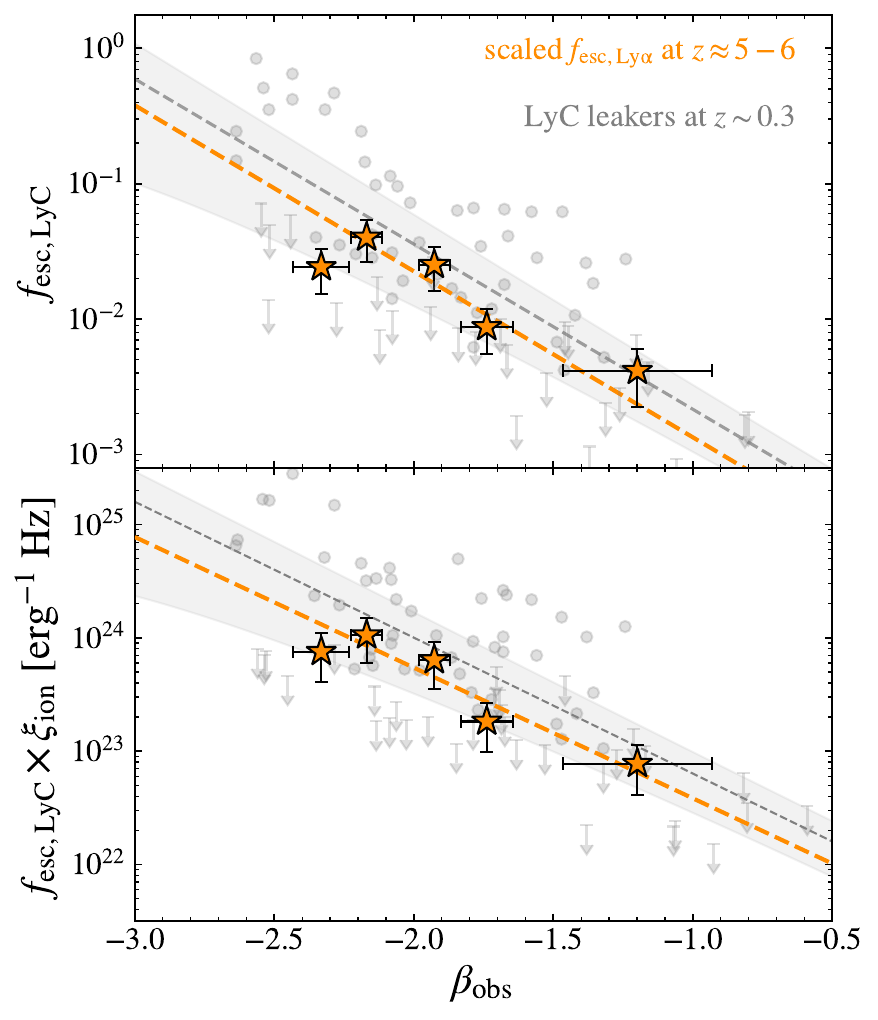}
    \caption{The relation between \betaobs\ and \fesclyc, \fesclyc\xiion. The gray circles represent LyC leakers at $z\sim 0.3$ from \cite{Chisholm2022}, where \betaobs\ are measured as observed stellar continuum slope at 1550\AA. The gray downward arrows refer to the $1\sigma$ upper limits of \fesclyc\ for galaxies without LyC detected, i.e. LyC flux is not significantly detected given the observed backgrounds. The gray dashed line and the shaded region are the analytical relations of \fesclyc\ and \fesclyc\xiion\ with respect to \betaobs, and the $1\sigma$ uncertainties of the relations. The orange stars are our stack \fesclya\ measurements assuming the \citetalias{Calzetti2000} dust law,  scaled by a factor of 0.15 according to \cite{Begley2023}, and the orange dashed lines are the best-fit \fesclya$-$\betaobs\ and \fesclya\xiion$-$\betaobs\  relation as described Equation \ref{Eq:fescLyA_EBV}, also scaled by 0.15.  }
    \label{fig:fescLyC_beta}
\end{figure}

In fact, the dependencies on \betaobs\ and \EBV\  are also found in \fesclyc\ of low-redshift LyC leaking galaxies.  With a compilation of LyC leakers at $z\sim0.3$, \cite{Chisholm2022} found an analytical relation between \fesclyc\ and \betaobs: $f_{\mathrm{esc}, \mathrm{LyC}}=(1.3 \pm 0.6) \times 10^{-4} \times 10^{(-1.22 \pm 0.1) \beta_{\mathrm{obs}}}$. The exponential index (-1.22) is in good agreement with that of our best-fit \fesclya$-$\betaobs\ relation (-1.23) in Equation \ref{Eq:fescLyA_EBV}.  Note that we only compare the \cite{Chisholm2022} \fesclyc$-$\betaobs\ relation with our  \citetalias{Calzetti2000} dust law-based \fesclya$-$\betaobs\ relation, because the \betaobs\ in \cite{Chisholm2022} are measured with the modeled stellar continuum assuming \cite{Reddy2016} attenuation law similar in shape to the \citetalias{Calzetti2000} law. The modeled \betaobs\ is sensitive to the effect of dust attenuation. Furthermore, \cite{Chisholm2022} models the SED of local LyC leakers using \textsc{starburst99} \citep{Leitherer1999}, while we perform SED modelling using \textsc{Beagle} which is based on \cite{BC03}; the different stellar continuum models also introduce systematics.  Nevertheless, the similarity in correlation trends in \fesclya$-$\betaobs\ and \fesclyc$-$\betaobs\ is consistent with a connection between the escape processes of \lya\ and LyC photons. Their escape processes are  sensitive to the opacity and geometry of neutral HI gas in both the vicinity of star-forming regions and the circumgalactic medium \citep{Dijkstra2016, Byrohl2020, Kimm2022, Maji2022}. Recently, \cite{Begley2023} investigated the LyC$-$\lya\ connection at $z\approx 4-5$ using \fesclyc\ predicted by the EWs of low-ionization, far-UV absorption lines in star-forming galaxies  \citep{Garilli2021} and IRAC-measured \ha. They derived a monotonic relation, \fesclyc$\simeq0.15_{-0.04}^{+0.06}$\fesclya. We scale our measured \fesclya\ and \fesclya$-$\betaobs\ relation by 0.15 in Figure \ref{fig:fescLyC_beta}. The scaled \fesclya$-$\betaobs\ relation is well consistent with the trend of LyC leakers within 1$\sigma$. Likewise, we estimate \fesclya\xiion\ as a function of \betaobs\  using the median stack \fesclya\ and the median values of \xiion\ in each stack bin, and then scale \fesclya\xiion\ by 0.15 as a proxy to \fesclyc\xiion, the production rate of ionizing photons that can escape to and ionize the IGM. As shown in Figure \ref{fig:fescLyC_beta}, the scaled \fesclya\xiion$-$\betaobs\ relation is also in good agreement with the \fesclyc\xiion$-$\betaobs\ relation in \cite{Chisholm2022}. It implies that \fesclya\xiion\ can be used to qualitatively estimate the total emitted ionizing emissivity that impacts the IGM (see \S\ref{sec:reionization} for more details)   Despite systematics, such as bias in sample selection and the slight difference in \lya\ transmission at $z=4-5$ and $z=5-6$,  the consistency of the scaled \fesclya\ and \fesclyc\ suggests that \fesclya\ is likely to be proportional to \fesclyc. With this scaling relation, the overall \fesclyc\ for our HAEs are $\sim 0.015$ (\fesclya$\approx$0.1) and thus negligible for the estimation of \xiion\ in Equation \ref{eq:Lha_fesc}.

The link between \lya\ and LyC escape is also in line with the correlations we observe between \fesclya\ and galaxy properties. If the physical conditions that favor LyC photon leakage also facilitate the escape of \lya,  observables that can indicate \fesclyc\ are expected to show strong correlations with \fesclya.   From cosmological radiation hydrodynamics simulations, \cite{Choustikov2023} concludes that a good diagnostic of \fesclyc\ should simultaneously encapsulate at least two of the three conditions: (1) high sSFR for the production of ionizing photons, (2) the age of stellar population and timescale for stellar feedback, and (3) a proxy for neutral gas content and the ionization state of ISM.  In this context, \betaobs\ and \EBV\  serve as strong indicators for the escape of LyC and \lya\ photons. \EBV\  directly tracks the dust content along the line of sight and may also trace neutral hydrogen \citep[e.g.,][]{Katz2022}; the production and destruction of dust are also related to supernovae activities \cite[e.g.,][]{Priestley2021}. Galaxies with bluer \betaobs\ are biased towards having higher sSFRs and younger stellar populations; \betaobs\ is also very sensitive to dust attenuation.  \EBV\  and \betaobs\ satisfy at least two of the three criteria and  thus are good probes of \fesclya.  The three physical conditions discussed above also help us to understand the relatively poor dependence of \fesclya\ on sSFR, \xiion,  \SDsSFR\ and \MUVatt. sSFR or \xiion\ represents just one facet of the three necessary conditions so that a high sSFR or \xiion\  only could not guarantee a high \fesclya.  \SDsSFR\ alone cannot be used to reveal feedback processes or the state of the ISM along the sightline. \MUVatt\ is a complex $M_\star$, dust, SFR, stellar population, etc., so the \fesclya-\MUVatt\ may suffer from large scatters introduced by the coupling of \MUVatt\ with respect to these parameters.  Also, \MUVatt\ do not trace the ISM state along the sightline. Although \cite{Choustikov2023} found \MUVatt\ can be a weak tracer of sSFR and dust attenuation at fixed $M_\star$, in which case \MUVatt\ satisfies two out of three conditions while the stellar age would still introduce some scatters, our sample size is not large enough to disentangle on the \MUVatt\ dependence of \fesclya\ at a fixed $M_\star$.

\subsection{Implications for cosmic Reionization}\label{sec:reionization}
There is an increasing body of evidences showing that dwarf galaxies are the main drivers of reionization \citep{Atek2023, Simmonds2023,Mascia2023}. However, the relative importance of massive and low-mass galaxies in
driving reionization remains uncertain \citep{Robertson2015, Finkelstein2019, Naidu2020}. It is necessary to quantify the relative contribution of galaxies with different UV magnitude or masses  to the  total ionizing emissivity, i.e., the number of ionizing photons emitted per unit time and comoving volume. $\dot{n}_{\rm ion}$ contributed by a specific population can be cast as a product of the total observed UV luminosity density $\rho_{\rm UV,obs}$, the production efficiency \xiion, and the escape fraction of ionizing photons \fesclyc:
\begin{equation}
 \dot{n}_{\text {ion }}=\rho_{\mathrm{UV, obs}} \times \xi_{\text {ion }} \times f_{\mathrm{esc}, \mathrm{LyC}}^{\rm rel}.
\end{equation}
where \fesclycrel\ is the escape fraction of LyC photons relative to the UV photons, defined as $f_{\mathrm{esc}, \mathrm{LyC}}^{\rm rel}=f_{\rm esc,LyC}/f_{\rm UV}=f_{\rm esc,LyC} 10^{0.4 A_{\rm UV}}$ \citep{Bouwens2016, Steidel2018, Bian2020}. We adopt $A_{\rm UV}$ from the best-fit SED models at the rest-frame 1500\AA. We emphasize that the definitions of ionizing efficiency and escape fraction of ionizing photons may vary among literature \citep{Bouwens2016,Steidel2018,Tang2019, Simmonds2023}, which may cause confusion.  In this section we follow the definition in \cite{Bouwens2016}. We define \xiion\ as the ionizing production rate per unit \textit{intrinsic} UV luminosity as described in \S\ref{sec:SED}. We adopt the relative escape fraction, \fesclycrel, so that the product \fesclycrel\xiion\ represents the production of ionizing photons that escape to and ionize the IGM per \textit{observed} UV luminosity density.  Likewise, we define the relative escape fraction of \lya\ photons as $f_{\mathrm{esc}, \mathrm{Ly\alpha}}^{\rm rel}=f_{\rm esc,Ly\alpha}/f_{\rm UV}=f_{\rm esc,Ly\alpha} 10^{0.4 A_{\rm UV}}$.
 
\begin{figure}
    \centering
    \includegraphics[width=\columnwidth]{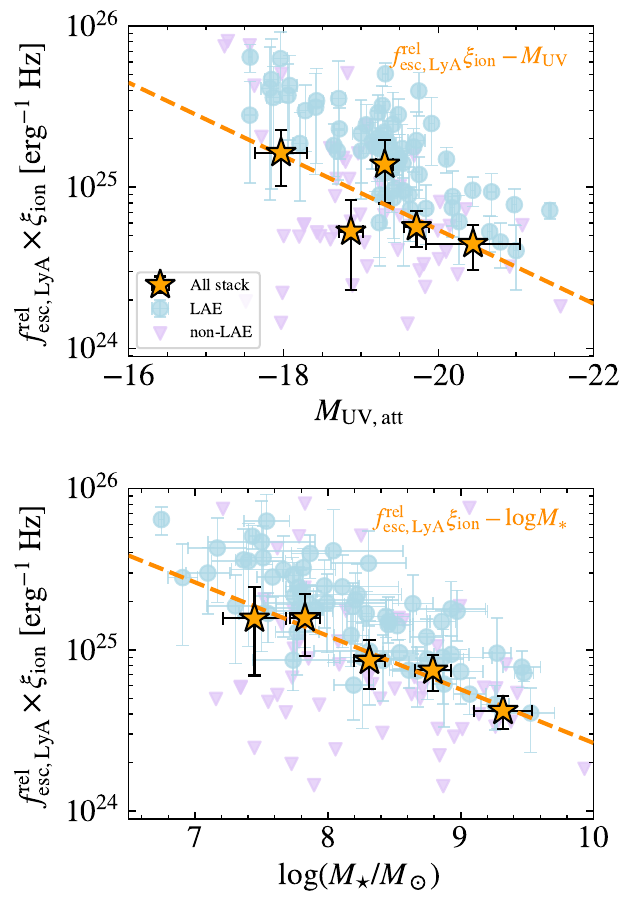}
    \caption{\fesclyarel\xiion\ as a function of   \MUVatt, and \Ms\  assuming the \citetalias{Calzetti2000} dust law. The blue circles indicate \fesclyarel\xiion\ for each individual LAEs and purple triangles are \fesclyarel\xiion\ of non-LAEs with forced measured \fesclyarel.  The fringed orange stars indicate the stack \fesclyarel\xiion\ for all HAEs. We adopt the median stack \fesclya\ values and the median \xiion\ and $A_{\rm UV}$ of HAEs in each stack bin.  We show the best-fit relations of the stack \fesclya\xiion\ of all HAEs with respect to \MUVatt, \Ms\ as the orange lines. }
    \label{fig:fescxiion_rel}
\end{figure}

If \fesclya\ is proportional to \fesclyc, as discussed in \S\ref{sec:lya and lyc}, \fesclyarel\xiion\ is proportional to \fesclycrel\xiion.   We investigate the dependence of \fesclyarel\xiion\  on \MUVatt\  and \Ms\ in Figure \ref{fig:fescxiion_rel}.  
We fit \fesclyarel\xiion\ as a function of  \MUVatt\ and \Ms, yielding:
\begin{equation}
\begin{split}
 \log f_{\rm esc, Ly\alpha}^{\rm rel}\xi_{\rm ion} =  (0.23\pm 0.10) & \times (M_{\rm UV,att}+19.5) \\
 &+ (24.84\pm 0.08)
  \end{split}
  \label{eq:fescxi_MUV}
 \end{equation}
\begin{equation}
\begin{split}
 \log f_{\rm esc, Ly\alpha}^{\rm rel}\xi_{\rm ion} = (-0.33\pm0.04) &\times  \log (M_{\star}/10^{8.5} M_\odot)   \\
 &+ (24.92\pm0.03)
  \end{split}
  \label{eq:fescxi_Ms}
 \end{equation}
To provide a reference for future works that may define the ionizing efficiency and escape fraction differently, we also present the analytical relations of \fesclya\xiion\ with respect to \MUVatt\ and \Ms\ in Appendix \ref{sec:abs_fescxiion_relation}, which can be proportional to the ionizing emissivity per unit \textit{intrinsic} UV luminosity \fesclyc\xiion.

\begin{figure}[htbp]
    \centering
    \includegraphics[width=0.9\columnwidth]{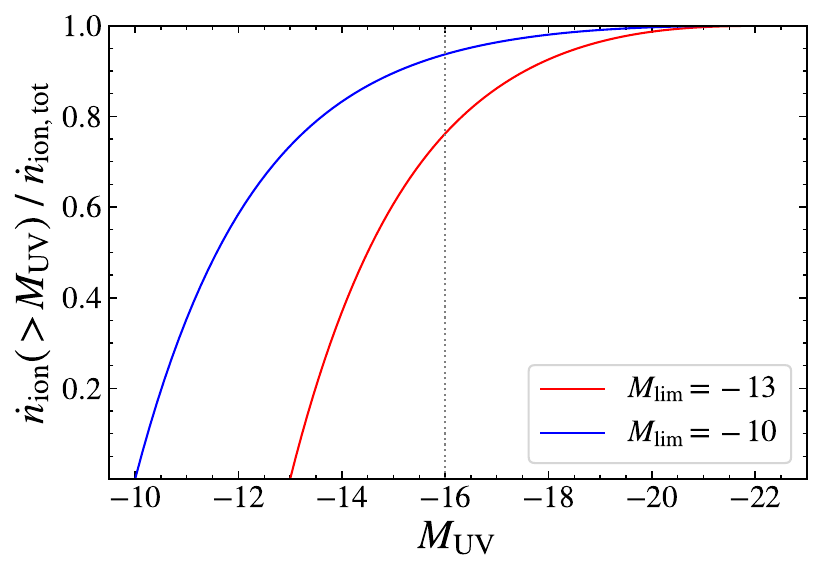}
    \caption{The accumulative emissivity of galaxies at $z=5.5$. Here $\frac{\dot{n}_{\rm ion}(>M_{\rm UV})}{\dot{n}_{\rm ion, tot}}$  refers to the fraction of ionizing budget contributed by galaxies fainter than a specific \MUVatt, as in Equation \ref{eq:nion}. We set the faint-end limit $M_{\rm lim}=-13$ and $-10$ for $\dot{n}_{\rm ion,tot}$, shown as the red and blue lines respectively.  We mark \MUVatt$=-16$ as the vertical dotted line  for reference. }
    \label{fig:accumulative_emissivity}
\end{figure}

The contribution to the total ionizing photon budget from galaxies fainter than $\rm M_{UV}$ can be expressed as:
\begin{equation}\label{eq:nion}
\begin{split}
\dot{n}_{\rm ion}(>{\rm M_{UV}}) \propto \int^{M_{\rm lim}}_{\rm M_{UV}} &f_{\rm esc, Ly\alpha}^{\rm rel} \xi_{\rm ion} (M) \\
& \times \Phi_{\rm UV}(M) L_{\rm UV}(M)dM
\end{split}
\end{equation}
where $M_{\rm lim}$ is the faint-end limit to compute the luminosity density by integrating over the UV LFs. We adopt the UV LF with a Schechter functional form and redshift-dependent Schechter parameters in \cite{Bouwens2022}. At $z=5.5$,  we have the characteristic luminosity $M^{*}=-21.02\pm0.04$,  the normalization $\Phi^{*}=0.56\pm0.04$ and the faint-end slope $\alpha=-1.90\pm0.02$. Although there are no direct observations on \fesclya\ of faint galaxies (\MUVatt$>-17$ mag), we can give a qualitative constraint on the contribution of UV faint galaxies by extrapolating the \fesclya\xiion$-$\MUVatt\ analytical relation  to the faint-end.  With $M_{\rm lim}=-13$ \citep{Bouwens2016} and Equation \ref{eq:fescxi_MUV},  we find that galaxies fainter than -16 mag will contribute about 72\% of the total ionizing budget, i.e., the accumulative emissivity fraction $\frac{\dot{n}_{\rm ion}(>-16\rm~ mag)} {\dot{n}_{\rm ion, tot} }\approx 0.76$ where $\dot{n}_{\rm ion, tot}$ is estimated by integrating from $-24$ mag to $M_{\rm lim}$. We present the  accumulative emissivity fraction as a function of \MUVatt\ in Figure \ref{fig:accumulative_emissivity}.  The fraction becomes $94\%$ if $M_{\rm lim}=-10$.  

Despite large uncertainties due to the lack of observations of \fesclya\ on the faint-end, our result suggests that UV-faint galaxies are the main contributor to the total ionizing emissivity, at least at $z=5-6$. If the \fesclya$-$\fesclyc\ connection and \fesclya\xiion\ dependence on \MUVatt\   hold at the epoch of reionization, UV faint galaxies will provide the majority of ionizing photons to reionize the Universe. \xj{These faint young galaxies may have elevated ionizing photon production efficiency \citep{Maseda2020}.} Further observations on \fesclya\ of faint galaxies (\MUVatt$>-17$ mag) are required to quantitatively constrain the ionizing emissivity from them. Moreover, to reveal the role of galaxies with different stellar masses during the reionization,  it is necessary to measure the stellar mass functions at high redshifts, and observe \lya\ emission of low-mass galaxies that extend Equation \ref{eq:fescxi_Ms} to the lower-mass end.

%% file: 06summary.tex
\section{Summary}\label{sec:summary}
We investigate the escape of \lya\ with a large sample of \nHAE\ HAEs at $z\approx 4.9-6.3$ in the GOODS-S field, which lies in the overlap region of the JWST FRESCO program and the public MUSE surveys. The JWST/NIRCam WFSS observations provided by FRESCO allow us to construct an unbiased HAE sample with precise measurements of their \ha\ emission lines. We search for their \lya\ emission in the 3D data cubes of VLT/MUSE.  We measure the \lya\ escape fraction (\fesclya) of individual galaxies and perform stacking analysis for the overall populations.  Our main conclusions are as follows:


\begin{itemize}
    \item[(\romannumeral1)] The overall \fesclya\ of HAEs at $z\approx5-6$ is \fesclyaCal\ (\fesclyaSMC) under the assumption of \citetalias{Calzetti2000} (SMC) dust attenuation law.  The distribution of \fesclya\ can be described using an exponential model with a median escape fraction of $0.10\pm0.01$ ($0.12^{+0.02}_{-0.01}$) for the \citetalias{Calzetti2000} (SMC) dust law.
    
    \item[(\romannumeral2)] \fesclya\ exhibits strong anti-correlations with the dust attenuation \EBV\  and the observed UV slope (\betaobs), indicative of the crucial role of dust plays in modulating the escape of \lya\ photons.  Adopting the \citetalias{Calzetti2000} (SMC) dust law, our best-fit relations between \fesclya\ and \EBV\  and \betaobs\ predict \fesclya$\approx0.48~ (0.46)$ for dust-free galaxies and \fesclya$\approx0.62~ (0.54)$ for young galaxies with very blue UV slopes \betaobs$\approx-2.5$.  

    \item[(\romannumeral3)] We find that \fesclya\ also correlates with the stellar mass (\Ms) and the intrinsic UV magnitude (\MUVint). The correlations are regulated by the underlying coupling between \Ms, \MUVint\  and \betaobs, \EBV\ .  We do not find strong correlations between \fesclya\ and the ionizing efficiency (\xiion), the observed UV magnitude (\MUVatt), the specific star formation rate  (sSFR) and the sSFR surface density (\SDsSFR). The variations of \fesclya\ are also regulated by the scatterings of \betaobs\ and \EBV\  coupled with these parameters. 

    \item[(\romannumeral4)] We find a tentative decrease in \fesclya\ as $z$ grows at $z\gtrsim5$ combined with measurements in \cite{Sun2022b}. It implies an increasing neutral fraction of IGM  at this epoch. At $z=5.2$ and $z=5.8$ our measured \fesclya\ are only $56\%$ and $37\%$ of the previous estimates based on \lya\ and UV luminosity functions \citep{Hayes2011}. The difference could arise from the uncertainties in estimating the \lya\ LFs and the conversion between UV and \ha\ luminosities in previous studies without direct \ha\ constraints.

    \item[(\romannumeral5)] The \fesclya$-$\betaobs\ and \fesclya\xiion$-$\betaobs\ relations has  log-space slopes very close to those of the \fesclyc$-$\betaobs\  and \fesclyc\xiion$-$\betaobs\ relations measured from LyC leakers at $z\sim0.3$ \citep{Chisholm2022}. The scaling factor is consistent with the monotonic \fesclya$-$\fesclyc\ connection at $z=4-5$ \citep{Begley2023}, implying that \fesclya\ can be proportional to \fesclyc. This may indicate that the escapes of \lya\  and LyC are regulated by similar physical processes. It is also in line with the weak dependence of \fesclya\ on \MUVatt,  \xiion, sSFR, and \SDsSFR. The four parameters are not enough to depict the physical conditions that favor LyC and \lya\ escape, as predicted by simulations \citep{Choustikov2023}, so they are all weak tracers of LyC or \lya\ in isolation.

    \item[(\romannumeral6)] 
    We define the relative escape fraction of  \lya\ photons \fesclyarel\ as \fesclya\ divided by the escape fraction of UV photons. We fit the analytical relations of \fesclyarel\xiion\ with respect to \MUVatt\ and \Ms, which is proportional to the ionizing emissivity per unit observed UV luminosity density. By extrapolating the \fesclyarel\xiion$-$\MUVatt\ relation to the faint end, we calculate the fraction of ionizing emissivity from  UV faint galaxies (\MUVatt$>-16$ mag).  
    Qualitatively, we infer that galaxies with UV fainter than $-16$ mag may contribute $>70\%$ of the ionizing budget at $z=5.5$.  If the dependence of \fesclya\xiion\ on \MUVatt\ holds till the epoch of reionization, UV-faint galaxies are predicted to be the main contributor to the reionization. Note that our HAE samples only reach \MUVatt$\sim-17$ mag, so direct constraints on \fesclya\ at the fainter end are required to better quantify the ionizing budget from faint galaxies. 
\end{itemize}


JWST has shown promising prospects to unveil the mysteries of the EoR.  Deeper and wider spectroscopic surveys in the future 
would reach the fainter end with lower stellar masses, and further probe the escape process of \lya\ and LyC photons in newborn infant galaxies.

%% file: 99DataAvail.tex
\section*{Data Availability}

The HAE catalog and stack results analyzed in this work are available on Zenodo: \dataset[doi:10.5281/zenodo.10802630]{https://doi.org/10.5281/zenodo.10802630}.  \bigskip

Some of the data presented in this paper were obtained from the Mikulski Archive for Space
Telescopes (MAST) at the Space Telescope Science Institute. The specific observations analyzed
can be accessed via the FRESCO High Level Science Product \citep{10.17909/gdyc-7g80}.

%% file: 99other.tex
\section*{Acknowledgments}

We thank the anonymous referee for their comments that helped improve this paper. We thank P. A. Oesch for very helpful discussions. 
 X.L., Z. C., Y.W., Z.L., \& M.L. are supported by the National Key R\&D Program of China (grant no.\ 2018YFA0404503), the National Science Foundation of China (grant no.\ 12073014), and the science research grants from the China Manned Space Project with No. CMS-CSST2021-A05.
F.S.\ acknowledges support from the NRAO Student Observing Support (SOS) award SOSPA7-022. 
F.S.\ 
 contract to the University of Arizona, NAS5-02105.

This work is based on observations made with the VLT/MUSE, NASA/ESA Hubble Space Telescope and  NASA/ESA/CSA James Webb Space Telescope. MUSE observations used in this work are taken by the MUSE-Wide Survey and the MUSE Hubble UDF Survey as parts of the MUSE Consortium. HST and JWST data were obtained from the Mikulski Archive for Space Telescopes at the Space Telescope Science Institute, which is operated by the Association of Universities for Research in Astronomy, Inc., under NASA contract NAS 5-03127 for JWST and NAS 5–26555 for HST. The JWST observations are associated with program GO-1895. The HST observations are taken from Hubble Legacy Field Hubble Data Release.  The authors are sincerely grateful to the FRESCO team for generously developing this program with a zero-exclusive-access period. 

The authors also acknowledge Zechang Sun for maintaining the high-performance computing platform for the high-redshift research group in Department of Astronomy, Tsinghua University.

\vspace{5mm}
Some of the data presented in this paper were obtained from the Mikulski Archive for Space
Telescopes (MAST) at the Space Telescope Science Institute. The specific observations analyzed
can be accessed via \dataset[doi:10.17909/gdyc-7g80]{https://archive.stsci.edu/hlsp/fresco}.

\facilities{JWST/NIRCam, ALMA, VLT/MUSE}
\software{ Grizli \citep[v1.8.2][]{grizli}
JWST calibration pipeline \citep[v1.6.2][]{Bushouse2022}
    BEAGLE \citep{Chevallard2106}
    EAZY \citep{EAZY} 
    EMCEE \citep{emcee}
    SciPy \citep{scipy}
    ZAP \citep{Soto2016}
}

%% file: 99appendix.tex
\counterwithin{figure}{section}

\section{Sample Selection}\label{sec:sample_selection_detail}

\begin{figure*}[htbp!]
    \centering
    \includegraphics[width=\textwidth]{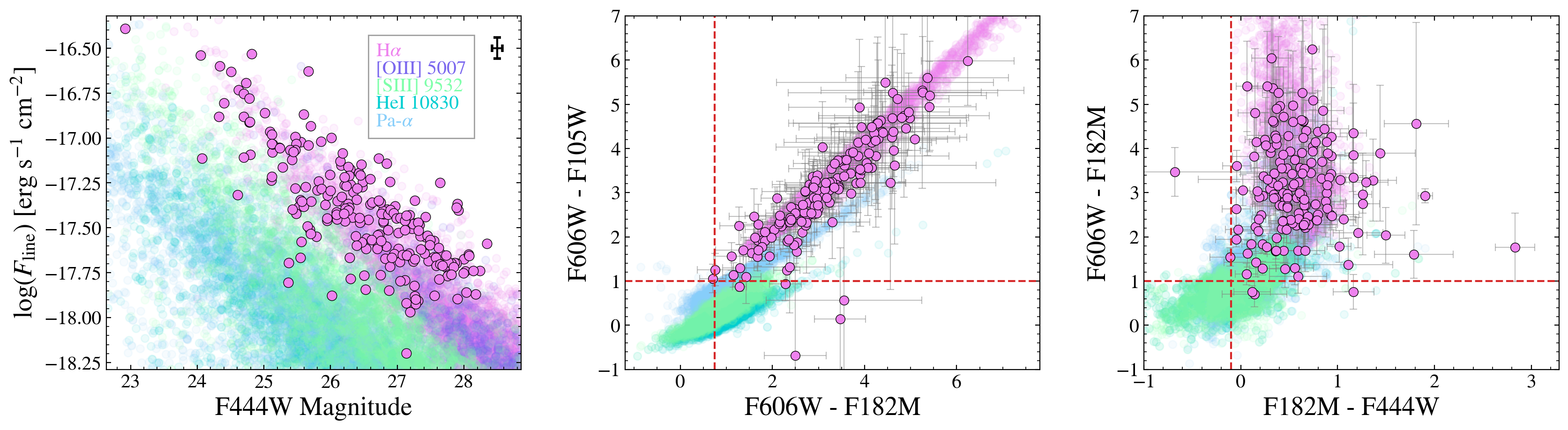}
    \caption{The flux distribution of selected HAEs (left panel) and color-color diagrams showing our photometric selection criteria (middle and right panel).  The magenta dots represent the \ha\ flux and magnitude of our HAE sample. In the left panel we label the typical uncertainty in line flux and magnitude. Overlying colored dots are emitters from \texttt{JAGUAR} mocks, residing in redshift windows that their specific emission lines could be captured by F444W grism: \ha\ at $z\approx 4.9-6.7$, [OIII]$\lambda5007$ at $z\approx6.9-9.0$, HeI at $z\approx2.5-3.6$, Pa$\alpha$ $\lambda10830$ at $z\approx 1.05-1.7$ and [SIII] $\lambda 9532$ at $z\approx 3.0-4.3$.  Our selection criteria matches well with mock HAEs and can effectively rule out most low-redshift objects.}
    \label{fig:photometry_selection}
\end{figure*}

We present the distribution of the magnitude versus line flux and the photometric criteria for our HAE selection in Figure \ref{fig:photometry_selection} as described in \S\ref{sec:selection}. We select mock galaxies from \texttt{JAGUAR} \citep{Williams2018} which have certain emission features captured by NIRCam/WFSS F444W grism, and model their lines with Gaussian profiles on the mock spectra.  To mimic the observational limits, we cut the mock galaxy sample by a magnitude limit of 29 for F444W and flux limit of $10^{-18.3}$ erg s$^{-1}$ cm$^{-2}$ for emission lines. As illustrated in  Figure \ref{fig:photometry_selection}, the colors of our sample are generally consistent with those  mock HAEs and could be distinguished effectively  from emitters at lower redshifts. 
In the left panel of Figure \ref{fig:photometry_selection}, these HAEs stay at a tightly linear region in the magnitude$-$\ha\ flux diagram, possibly suggesting main sequence galaxies at this redshift range. They perfectly overlap with the mock HAEs, and could be distinguished easily from emitters at lower redshifts. The main possible contamination is [OIII] emitters at higher redshifts, most of which could be successfully picked out by the procedures in \S\ref{sec:final_ha_sample}. 

\section{\fesclya\ under the assumption of SMC dust attenuation law}

\begin{figure}
    \centering
    \includegraphics[width=\columnwidth]{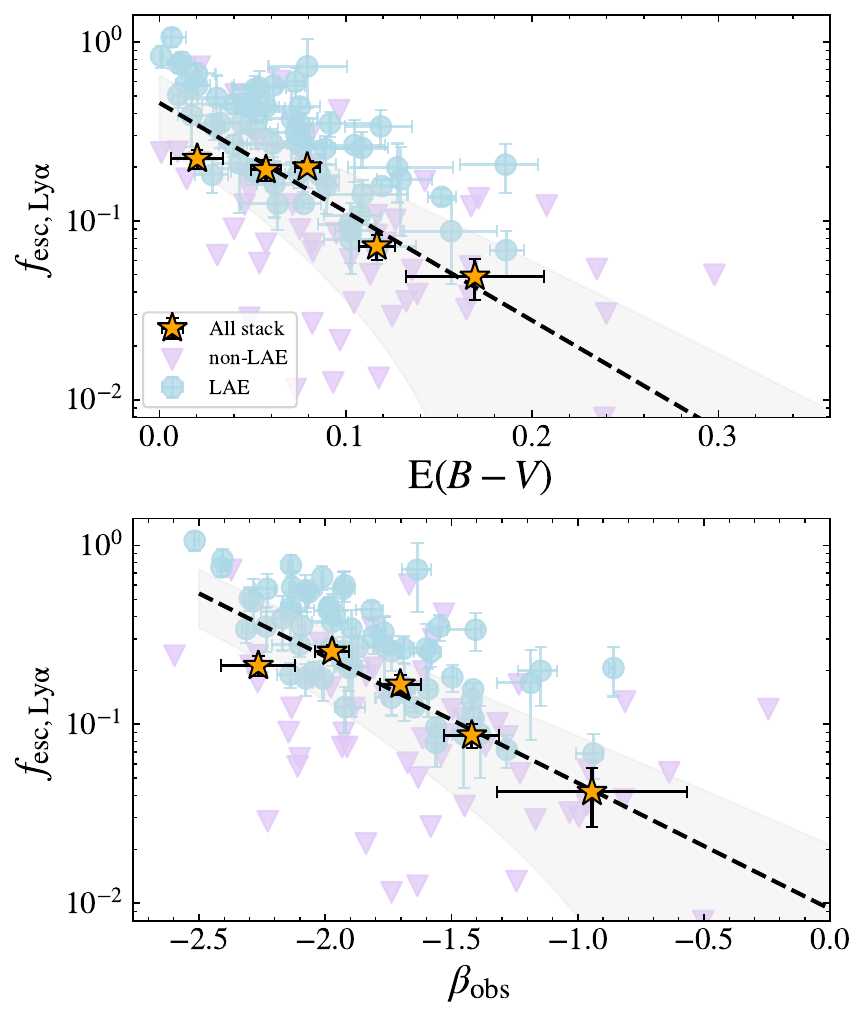}
    \caption{Same as Figure \ref{fig:fesclya_dust} while assuming the SMC dust attenuation law. The dahsed lines and gray shaded regions represent the best-fit analytical relation as described in Equation \ref{Eq:fescLyA_EBV}.}
    \label{fig:fescLyA_EBV_beta_obs_SMC}
\end{figure}

As discussed in \S\ref{sec:SED}, the SMC dust attenuation law is broadly adopted in high-redshift studies.  It is suggested by previous sub-millimeter observations \citep[e.g.,][]{Capak2015} for typical $
z\sim5-6$ galaxies. Conventionally, \citetalias{Calzetti2000} and SMC attenuation laws are both applied and complement each other \citep[e.g.][]{Bouwens2016,Shivaei2018}. Figure \ref{fig:fescLyA_EBV_beta_obs_SMC} illustrates \fesclya\ of individual galaxies and from median-stack measurements based on the SMC dust law. 

\section{The analytical relations of \fesclya\xiion\ with respect to \betaobs, \MUVatt, and \Ms}\label{sec:abs_fescxiion_relation}

\begin{figure*}
    \centering
    \includegraphics[width=\textwidth]{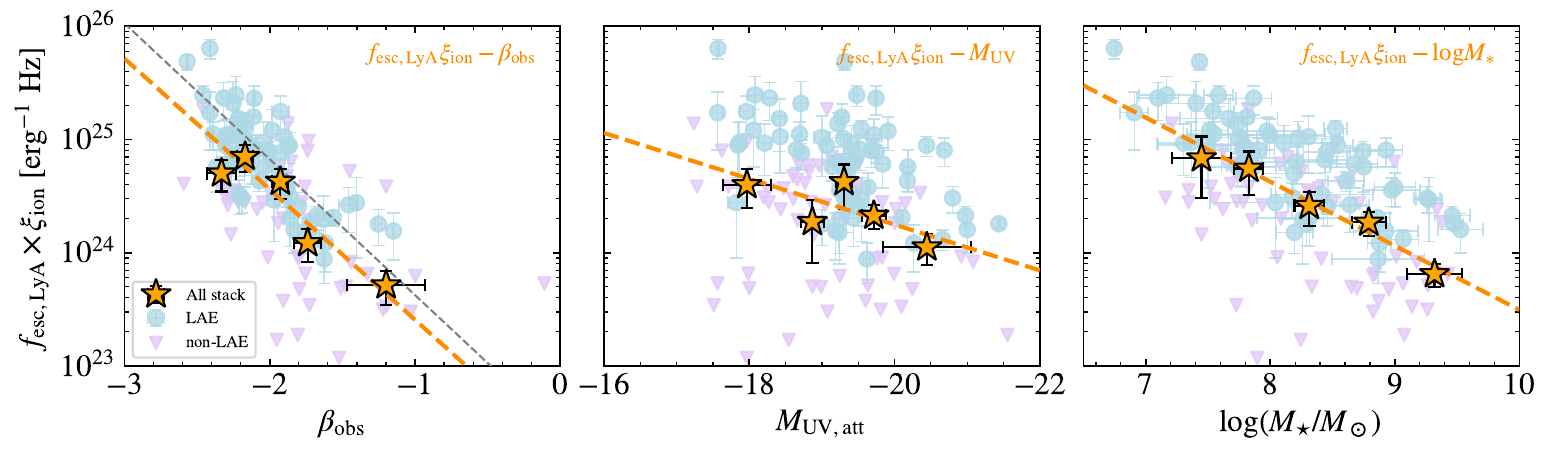}
    \caption{ \fesclya\xiion\ as a function of \betaobs, \MUVatt, and \Ms\  assuming the \citetalias{Calzetti2000} dust law. The blue circles indicate \fesclya\xiion\ for each individual LAEs and purple triangles are \fesclya\xiion\ of non-LAEs with  forced measured \fesclya.  The fringed orange stars indicate the stack \fesclya\xiion\ for all HAEs. We adopt the stack \fesclya\ values and the median \xiion\ of HAEs in each stack bin.  We show the best-fit relations between the stack \fesclya\xiion\ of all HAEs and \betaobs, \MUVatt, \Ms\ as the orange lines.  In the left panel, we show the scaled \fesclyc\xiion$-$\betaobs\ relations, which is the \fesclyc\xiion$-$\betaobs\ relation in \cite{Chisholm2022} divided by 0.15, as the gray dashed line.}
    \label{fig:fescxiion}
\end{figure*}

The definitions of ionizing efficiency, \xiion, may vary among the literature, from ionizing photon production rate per unit \textit{intrinsic} UV luminosity \citep[e.g.,][]{Bouwens2016, Chisholm2022} to ionizing photon production rate per unit \textit{observed} UV luminosity \citep[e.g.][]{Simmonds2023, Ning2023}. If \xiion\ is defined as the production rate per unit \textit{observed} UV luminosity, the absolute escape fraction, \fesclyc, should be used in Equation \ref{eq:nion} to calculate the ionizing emissivity per \textit{observed} UV luminosity density $n_{\rm ion}/\rho_{\rm UV,obs}$.  To provide a reference to future works that may adopt different definitions from this work, we also fit the analytical relations of \fesclya\xiion\ with respect to \MUVatt\ and \Ms, as supplements to Equation  \ref{eq:fescxi_MUV} and \ref{eq:fescxi_Ms}. We also presents \fesclya\xiion\ as a function of \betaobs, which is proportional to \fesclyc\xiion\ as shown in Figure \ref{fig:fescLyC_beta}.  As shown in Figure \ref{fig:fescxiion}, these yield

\begin{equation}
\begin{split}
     \log f_{\rm esc, Ly\alpha}\xi_{\rm ion} =  (-1.15\pm 0.35) &\times (\beta_{\rm obs} + 2.0) \\
     &+ (24.56\pm 0.09),
\end{split}
\label{eq:abs_fescxi_beta}
\end{equation}
\begin{equation}
\begin{split}
 \log f_{\rm esc, Ly\alpha}\xi_{\rm ion} =  (0.20\pm 0.09) & \times (M_{\rm UV,att}+19.5) \\
 &+ (24.35\pm 0.07),
  \end{split}
  \label{eq:abs_fescxi_MUV}
 \end{equation}
\begin{equation}
\begin{split}
 \log f_{\rm esc, Ly\alpha}\xi_{\rm ion} = (-0.57\pm0.07) &\times  \log (M_{\star}/10^{8.5} M_\odot)   \\
 &+ (24.63\pm0.05).
  \end{split}
  \label{eq:abs_fescxi_Ms}
 \end{equation}

\section{Double-peaked LAEs}\label{sec:double_peak_LAE}

\begin{figure}
    \hspace{3cm}
    \centering
   \includegraphics[width=1\columnwidth]{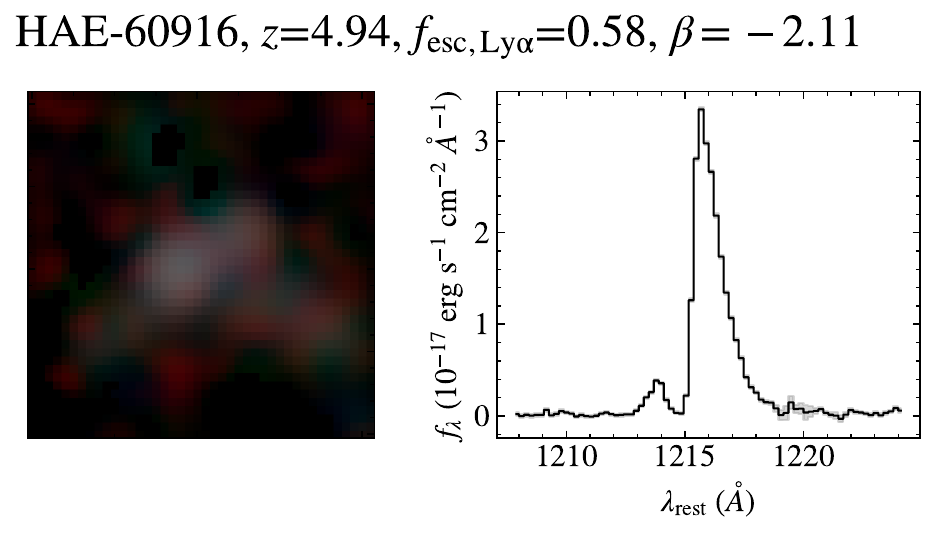}
 \includegraphics[width=1\columnwidth]{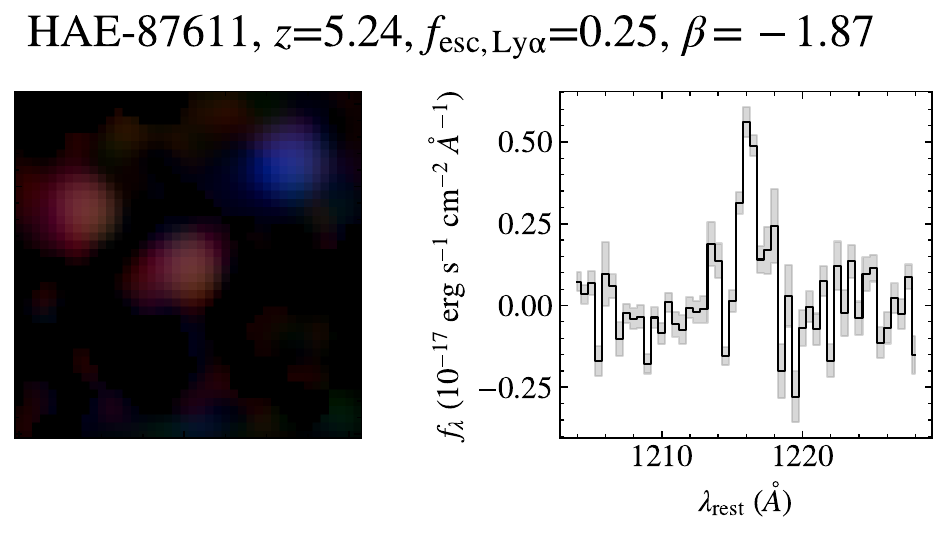}
    \caption{The two double-peaked \lya\ emitting galaxies in our final HAE sample. For each source, we show the $1.25\arcsec\times 1.25\arcsec$ RGB thumbnail in the right panel, composed of JWST F182M, F210M and F444W. The left panels show the rest-frame \lya\ spectra extracted from the MUSE cubes. }
    \label{fig:double_peak_LAE}
\end{figure}

\xj{As discussed in \S\ref{sec:fesclya_galaxy_properties}, among the 165 HAEs we find two LAEs presenting double-peaked \lya\ profiles, shown in Figure \ref{fig:double_peak_LAE}.  Their \fesclya\  ($0.58\pm0.11$ for HAE-60916 and $0.25\pm0.04$ for HAE-87611) are $2\sigma$ higher than the predicted values based on their \betaobs\ using Equation \ref{Eq:fescLyA_EBV} (predicted $0.20\pm0.14$ for HAE-60916 and $0.10\pm0.09$ for HAE-87611). The blue \lya\ peaks of the two sources are both weaker than the red peaks. Considering the strong attenuation of the IGM, the intrinsic emissivity of the blue peaks can be much stronger and possibly stronger than the red peaks.  The presence of blue \lya\ peak at $z>5$ might be a signal of enhanced escape of \lya\ photons compared to normal star-forming galaxies. This is consistent with the scenarios of double-peaked LAEs with strong blue peaks at cosmic noons  \citep{Furtak2022, Marques2022}. They generally present very blue \betaobs, high-ionization state (e.g., high [\ion{O}{3}/\ion{O}{2}]) and potential high \lya\ or LyC escape fraction, and might be experiencing an extreme star formation episode triggered by gas inflows.}

\section{Possible systematics for \lya\ flux measurement}

In this section, we discuss possible systematics and the uncertainties it may introduce during the measurement of \lya\ flux. We emphasize that, the systematics we mention in the following can only introduce negligible uncertainties in the final \fesclya\ estimates, proved by our well-tried experiments.

\subsection{Contamination rate}
We compare our HAE catalog with the public JADES-Deep catalog, which contains not only photometric redshifts for each source based on JADES' deep imaging but also spectroscopical confirmation for some of the sources combined with FRESCO's WFSS data, identified by the JADES team independently  (Sun in prep).  There are 80 HAEs in our sample covered by JADES-Deep, among which $\gtrsim$91\% have photo-$z$ in the range of 4.9--6.7, predicting \ha\ in F444W.  Further, $\gtrsim$70\% of them are reported to have \ha\ emission in JADES's catalog. These values are simply lower limits because we haven't considered the mismatch due to different segmentation strategies during the catalog matching. Indeed we do check the sources without JADES confirmed spec-$z$, and find most of them are (1) multi-components and suffering mismatch issues, (2) missed by the JADES team as they only report \ha\ emission with SNRs$\geq5$ (chat with Sun in private), while our detection threshold is SNRs$\geq4$. To sum up, our HAE catalog is reliable with high purity.

\subsection{The accuracy of \lya\ measurement}\label{sec:accuracy test}

As described in \S\ref{sec:MUSE_lya}, \cite{Kerutt2022} and \cite{Bacon2023} release their LAE catalog utilizing MUSE-Wide and MUSE-Deep survey data respectively. There are 39 isolated HAEs recorded in the two literature catalogs. Different extraction and flux measurement methods are adopted, including blind line searching algorithm \textsc{LSDCat} \citep{Herenz2017} and HST-prior extraction \textsc{ORIGIN} \citep{Mary2020}. In this work, instead, we extract the spectra from the exposure-time-weighted coadded datacube using circular and boxy apertures, and measure the line flux by fitting skewed Gaussian profiles.  For $\rm EW_{Ly\alpha}$, \cite{Kerutt2022} takes a fixed UV slope, $\beta=-1.97$, for their entire LAE sample and extrapolates the measured flux at the effective wavelength of the HST bands for the continua at the \lya\ wavelength, while \cite{Bacon2023} adopts the values provided by \textsc{pyPlatefit}\footnote{\url{https://github.com/musevlt/pyplatefit}}. Their methods are totally different from ours, which yields the continuum flux density for \lya\ lines by extrapolating the best-fit SED models. We compare the reported \lya\ properties of the 39 HAEs from the two catalogs with our own measurements, yielding a median flux ratio of $1.04\pm0.38$ and ${\rm EW}$ ratio of $0.85\pm0.18$.  Though the scattering is large, especially at the faint \lya\ end, our measurement is consistent with the literature within $1\sigma$. 
 

\subsection{Stack of median-filtered datacubes with different depths}
The MUSE datacubes used in this work have very different depths, ranging from 1h for MUSE-Wide to $\gtrsim$140h for MXDF.  Therefore the  \lya\ SNR for each individual HAE could vary dramatically. Theoretically, their median-stack is expected not to be biased towards those with highest SNRs, i.e. those covered by MUSE-Deep. Another concern is the median filtering we use to remove the continuum and nearby sources. It could possibly reduce the \lya\ flux by suppressing the wing of the emission lines, even though the optimized window size is chosen. We perform a test to check if these effects would significantly change our result. 

Our goal is to check whether the median stack of median-filtered spectra can recover the intrinsic flux. To start with, we generate a series of mock 1D spectra in a spectral sampling of 0.5\AA, with random SNRs and continua but uniform flux. 70\% spectra have low SNRs uniformly distributed from 0.1 to 3, 15\% have intermediate SNRs from 3 to 5, and the remaining have the highest SNRs from 5 to 15. We then remove the continua using a median filter of 101 pixels, the same as used in the real MUSE data, and stack all the mock spectra by taking the median value at each wavelength grid. We perform experiments for 500 times using 20, 30, 100, 200 spectra. The median values of the recovered flux for each setting reach 96--99\% of the initially set value, suggesting that a $\lesssim$5\% underestimate is possible. For a \fesclya\  $\sim$0.1, this systematics would simply result in an error of $\lesssim$0.005, smaller than the uncertainty as we report in this work. 